\documentclass[numbers,final,3p,times]{elsarticle}
\usepackage{amsthm}
\usepackage{array}
\usepackage{ulem}
\usepackage{amsmath}
\usepackage{amssymb}
\usepackage{bm}
\usepackage{CJK}
\usepackage{indentfirst}
\usepackage{mathrsfs}
\usepackage[raggedright]{titlesec}
\usepackage{verbatim}
\usepackage{color}
\usepackage{pdflscape}
\usepackage{threeparttable}
\usepackage{longtable} %±í¸ñ
\usepackage{threeparttable} %±í¸ñÄÚ¼Ó×¢ÊÍÓë±í¸ñÍ¬¿í
\usepackage{subfigure} %×ÓÍ¼
\usepackage{ulem} %É¾³ýÏß
\addtocounter{MaxMatrixCols}{50}

\allowdisplaybreaks[4]
\begin{document}
%\begin{spacing}{1.0}
\journal{}

%\begin{CJK*}{GBK}{song}
\begin{frontmatter}
%%\newcommand{\hei}{\CJKfamily{hei}}

%%\titlespacing*{\section}{10pt}{2ex plus .3ex minus .2ex}{-.2ex plus .2ex}

%%\titleformat*{\section}{\raggedright\bf\zihao{1}}
%%\titlespacing*{\subsection}{10pt}{1ex plus .3ex minus .2ex}{-.2ex plus .2ex}

%\title{Mesoscience-based structural theory for the spatiotemporal dynamics of heterogeneous gas-solid flows}
\title{Mesoscience-based structural theory for heterogeneous gas-solid flows}
\author[label1,label2]{Yige Liu}
\author[label1,label3]{Bidan Zhao}
\author[label1,label3]{Junwu Wang\corref{cor1}}
\cortext[cor1]{Corresponding author}
\ead{jwwang@cup.edu.cn}
\address[label1]{State Key Laboratory of Mesoscience and Engineering, Institute of Process Engineering, Chinese Academy of Sciences, P. O. Box 353, Beijing 100190, P. R. China}
\address[label2]{School of Chemical Engineering, University of Chinese Academy of Sciences, Beijing, 100049, P. R. China}
\address[label3]{Beijing Key Laboratory of Process Fluid Filtration and Separation, College of Mechanical and Transportation Engineering,
China University of Petroleum-Beijing, Beijing 102249, China}
%%\date{}
%%\maketitle

\begin{abstract}
Quantification of the spatiotemporal dynamics of heterogeneous gas-solid flows is critical for the design, scale-up and optimization of gas-solid reactors. In this article, by using the core concept of mesoscience (the compromise in competition between dominant mechanisms), a mathematically rigorous procedure is proposed to develop a mesoscience-based structural theory for the dynamics of heterogeneous gas-solid flows via describing the physical states corresponding to the realization of dominant mechanisms as the interpenetrating continua, finding the microscale governing equations of gas-solid flow, defining a dominant mechanism indicator function, and finally, performing ensemble averaging to obtain the macroscopic governing equations. It is shown that the theory can be mathematically formulated as partial differential equations (PDE) constrained dynamic optimization, possible closures for the constitutive relationships are then discussed, and numerical validations using a simplified theory are also performed. Present study (i) offers an alternative method to the popular two-fluid model for simulating the spatiotemporal dynamics of heterogeneous gas-solid flows, with the distinct advantage that the constitutive relationships can be developed using the models and correlations that are obtained from homogeneous systems, of course an extra model for the interphase mass transfer rate between dilute phase and dense phase is needed, and the effective numerical solution of PDE-constrained dynamic optimization needs to be explored; and (ii) provides a feasible procedure to formulate concrete mathematical equations from the core concept of mesoscience.
\end{abstract}


\end{frontmatter}

\section{Introduction}
Complex gas-solid two-phase flows are widely encountered in nature and industry \cite{kunii1991fluidization}. Deep understanding and mastering the hydrodynamics of gas-solid flows are critical for the efficient, sustainable and green development of chemical industry \cite{kunii1991fluidization,yang2020progress,zou2020laboratory}. However, gas-solid flows are notoriously difficult to theoretical analysis, experimental measurement and numerical modeling, due to their inherent essence of multiple spatiotemporal scales, the spatiotemporal evolution of mesoscale structures and its strong coupling with the mass-heat-energy transfer and chemical reactions \citep{li2013multiscale,wang2020continuum}. Therefore, our understanding is still far from perfect after century-long researches, reactor design, scale-up and optimization remain depending mainly on experiences \citep{li2013multiscale}. Nevertheless, many multiphase computational fluid dynamics (CFD) methods have been proposed and developed \cite{gidaspow1994multiphase,enwald1996eulerian,jackson2000dynamics,deen2007review,zhu2007discrete,van2008numerical,wang2009review,fox2012large,deen2014review,tenneti2014particle,
ge2017discrete,sundaresan2018toward,ge2019multiscale,wang2020continuum,alobaid2022progress,xu2022discrete}.

Two-fluid models and two-phase models have been developed in parallel, but toward completely distinct directions. Two-fluid model (TFM) \citep{anderson1967fluid,gidaspow1994multiphase} treats the gas and particles as two interpenetrating continua, various averaging techniques are then used to develop the governing equations that are based on the conservation principle of mass, momentum and energy. TFM targets on the quantification of spatiotemporal evolution of gas-solid flows from the very beginning of model development, but the constitutive relationships are established on the basis of (nearly) homogeneous structures inside each fluid element or computational grid. Therefore, TFM-based direct numerical simulation (TFM-based DNS) is needed to explicitly resolve the effects of mesoscale structures \citep{agrawal2001role,wang2009two,wang2010cfd}, or else, mesoscale-structure-dependent constitutive laws need to be developed properly \citep{xu1998multi,buyevich1999particulate,agrawal2001role,xiao2003theoretical,yang2003cfd,wang2006multi,wang2007simulation,igci2008filtered,wang2008eulerian,shi2011bubble,shuai2012cluster,lv2014simulation,
wang2016toward,schneiderbauer2017spatially,qin2019emms,he2020unified,hu2020cfd,jiang2021development,ouyang2021data,zhao2021multiscale,li2022dynamic,rauchenzauner2022dynamic}.
Details can be found in recent reviews \citep{wang2009review,wang2020continuum}.
On the other hand, two-phase models \citep{tommey1952gaseous,davidson1963fluidized,grace1974two,li1988method,li1994particle}, such as bubble-emulsion model for bubbling fluidized beds, gas-cluster model and energy-minimization multi-scale (EMMS) model for circulating fluidized bed risers, are established directly focusing on the quantification of two-phase structures. But they are usually formulated as global models for the hydrodynamics of whole reactor, therefore, they are typically unable to describe the spatiotemporal evolution of gas-solid flows.

In view of the cons and pros of two-fluid models and two-phase models, ideas have been inspired to combine them to develop two-fluid model with structural effect, which uses the merits of the strict fulfilment of conservation principle and transient nature of two-fluid model, and the (extended or modified) two-phase models for developing mesoscale-structure-dependent constitutive relationships.
This idea results in for example EMMS-based two-fluid model \citep{xiao2003theoretical,yang2003cfd,wang2007simulation,wang2008eulerian,nikolopoulos2010advanced,shi2011bubble,hong2012emms,wang2012emms,schneiderbauer2014filtered,lu2019energy,
qin2019emms,he2020unified,hu2020cfd,yang2021coupling,zhao2021multiscale}, cluster structure-dependent model \citep{shuai2011modeling,shuai2012cluster,xiaoxue2020comparative,li2022dynamic,sun2022numerical}, and bubble structure-dependent model \citep{lv2014simulation,wang2014modeling,li2017cfd,du2019numerical,zou2019cfd,zhao2021cfd}. More details can be found in recent reviews \citep{wang2009review,wang2020continuum}.

In this article, an attempt is made to keep the advantages of two-phase models for quantifying the effects of mesoscale structures but to remedy the disadvantages of the lack of strict mass, momentum and energy conservation as well as the short of transient dynamics, by learning the strengths of two-fluid models. More specifically, the core concept of mesoscience \citep{li2013multiscale,li2018multiscale,li2018mesoscience,chen2022multilevel}, the compromise in competition between dominant mechanisms, is used to develop a mesoscience-based structural theory for the spatiotemporal dynamics of heterogeneous gas-solid flows. The remaining of the article is organized as follows: Section 2 provides a detailed elaboration of the idea behind the mesoscience-based structural theory as well as a rigorous mathematical derivation of the theory; Section 3 provides a validation of the theory using a simplified model where the effects of key model input (the cluster size) are also studied; Finally, conclusions are made in Section 4.

\section{Mesoscience-based structural theory}
The study of mesoscience has established the core concept of compromise in competition between dominant mechanisms for describing the behavior of complex systems that are far away from thermodynamic equilibrium \citep{li2013multiscale,li2016focusing,huang2018mesoscience,li2018multiscale,li2018mesoscience,chen2022multilevel}, which results in the formulation of stability condition as its unique feature. However, a rigorous mathematical theory using the concept of compromise in competition between dominant mechanisms to describe the spatiotemporal dynamics of complex systems is not available yet, although a static model (the energy minimization multi-scale (EMMS) model) has been established to quantify the behavior at the nonequilibrium steady state \citep{li1988method,li1994particle}.
Present study attempts to extend those studies to develop a mesoscience-based structural theory for the spatiotemporal dynamics of heterogeneous gas-solid flow. As will be seen, the theory developed here may be regarded as an extension of the static and global EMMS model \citep{li1988method,li1994particle} to a transient and local model that is able to describe the spatiotemporal dynamics of heterogeneous gas-solid flows with rigorous mass, momentum and energy conservation, and may also be regarded as a much more rigorous formulation of EMMS-based two-fluid model using mesoscience \citep{wang2012emms}.

The structure of this Section is organized as follows: Section 2.1 introduce the idea behind the mesoscience-based structural theory; Section 2.2. derive the microscopic governing equations for homogeneous gas-solid mixtures within the dilute and dense phases; In Section 2.3, ensemble-average of microscopic governing equations is then carried out using the dominant mechanism indicator function as defined in Equation (\ref{C1}), and the stability condition of the system is analyzed. With those two steps, the framework of the mesoscience-based structural theory are formulated; Possible constitutive relationships are then discussed in Section 2.4.

\subsection{Idea of mesoscience-based structural theory}
Gas-solid flow is a typical dissipative system \citep{li1996gas} and the system is finally maintained at a nonequilibrium steady state \citep{wang2020continuum}.
In such a complex system, the gas-solid flow is characterized by the multiscale structures and their dynamic evolution \citep{li2003exploring}, where the collective behavior of massive numbers of particles is critical for quantifying the transport and reactive properties, and entirely new physical laws may appear at mesoscale. The multiscale structures are manifested by particle-scale or microscale structures, mesoscale bubbles and/or clusters structures and reactor-scale or macroscale structures in gas-fluidized beds \citep{li1994particle}. Specifically, the mesoscale structures, which is the focus of present study, feature the coexistence of gas-rich dilute phase and particle-rich dense phase in fluidized beds and the coexistence of bubble phase and emulsion phase in bubbling fluidized beds. For convenience of statement, we simply use dilute phase and dense phase to describe the two-phase structures, where the dilute phase can represent either bubble phase or gas-rich dilute phase, and the dense phase represents either emulsion phase or particle-rich dense phase. Furthermore, the study of gas-solid and other complex systems does reveal the existence of a new, general physical law at mesoscale, that is the compromise in competition between dominant mechanisms \citep{ge2007analytical}. This finding enables the appearance of mesoscience \citep{li2013multiscale,li2016focusing,huang2018mesoscience,li2018multiscale,li2018mesoscience,chen2022multilevel}.

Mesoscience has concluded that (i) there are at least two dominant mechanisms in a complex system, where each dominant mechanism corresponds to an extremum tendency representing by the realization of a physical state having similar dynamics, for example, the gas-solid flow within the dilute phase or the dense phase is similar. The compromise in competition between dominant mechanisms results in the multiscale structures featuring intrinsic correlation, strong spatiotemporal coupling and interactions; (ii) the relative importance of the dominant mechanisms results in the diversity of multiscale structures of complex systems; (iii) the nonequilibrium steady state of a complex system is constrained by the stability condition which represents the compromise in competition between dominant mechanisms, therefore, stability condition is an indispensable part of the physical theory that describes the behavior of complex systems, in addition to the constrained conditions such as the conservation equations.

\begin{figure}
\centering
\includegraphics[width=1.0\textwidth]{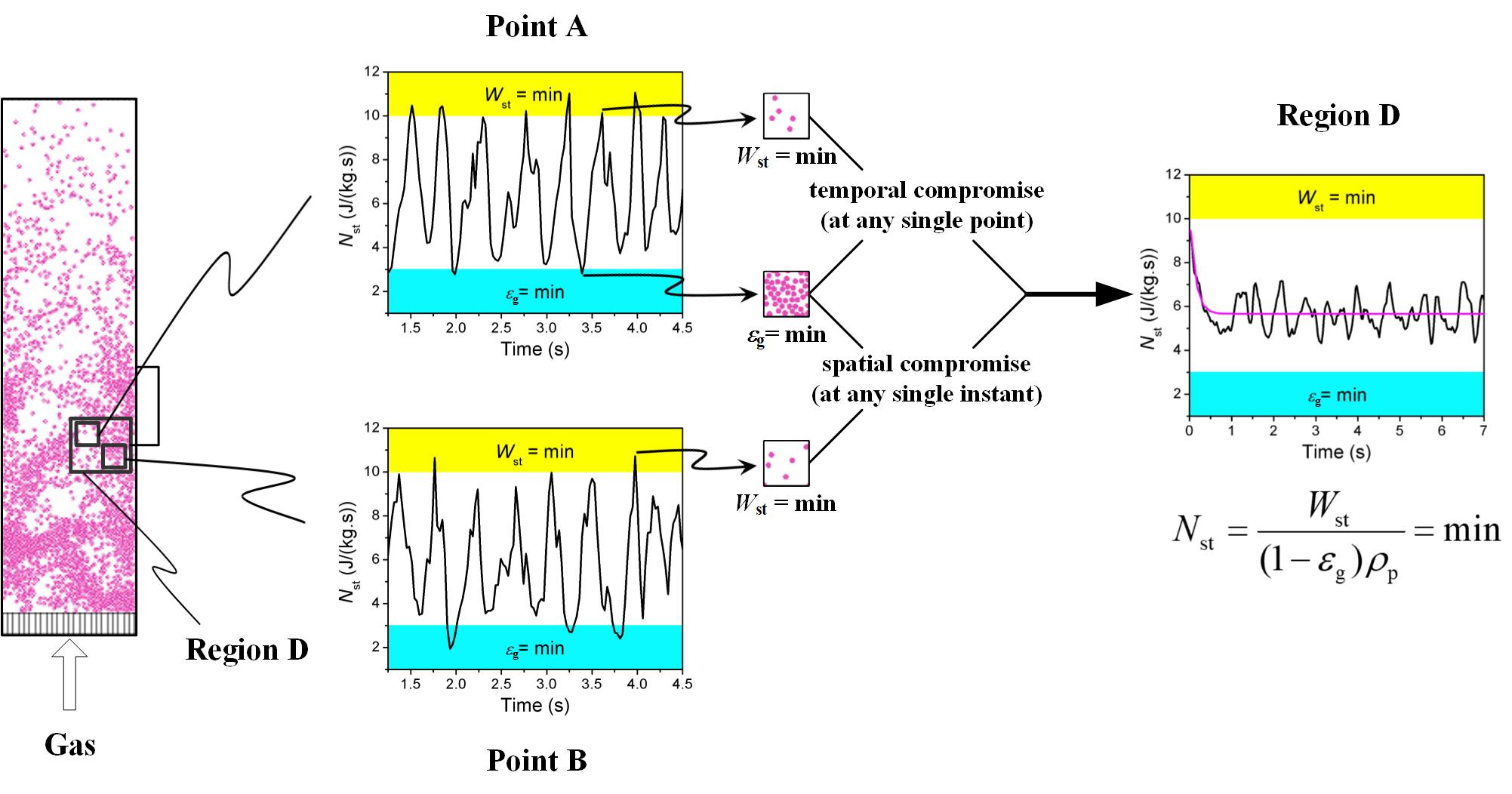}
\caption{Spatiotemporal compromise between dominant mechanisms and the behavior of the stability criterion of gas-solid flows at different scales, modified with permission from \cite{li2004multi}. Copyright 2004 Elsevier.}
\label{FigproofEMMS}
\end{figure}

To develop the mesoscience-based structural theory using the concept of compromise in competition between dominant mechanisms,
we start from the analysis of the behavior of stability criterion of gas-solid flow at different scales and the proof of the existence of stability condition in gas-solid flow \citep{li2004multi,zhang2005simulation}.
For the complex gas-solid system studied here, there are the gas-dominant mechanism and the particle-dominant mechanism. When the system is fully dominated by gas (i.e. the gas-dominant mechanism is fully fulfilled), the extremum tendency or the stability condition for the system is the minimization of of volume-specific energy consumption rate for transporting and suspending particles $W_{st}=min$, manifesting a dilute phase. On the other hand, the extremum tendency is the minimization of the voidage or the gas phase volume fraction $\varepsilon_g$ when the system is fully dominated by particles, i.e. $\varepsilon_g=1-\varepsilon_s=min$, manifesting a dense phase.
As shown in Figure \ref{FigproofEMMS} \citep{li2004multi,zhang2005simulation}: it was found that at microscale or a local point (point A or point B), a dominant mechanism can fully fulfil its extremum tendency locally and instantaneously, but the gas-dominant mechanism and the particle-dominant mechanism can only be completely fulfilled alternately at different spatial locations and different temporal locations (there is no room for the simultaneous realization of the two mechanisms at a same microscale spatiotemporal location). Those facts are indicated by the temporal fluctuations of the mass-specific energy consumption rate that is defined as $N_{st}=\frac{W_{st}}{(1-\varepsilon_g)\rho_p}$ at point A and point B without extremum tendency, where $\rho_p$ is the density of particles. The spatial and/or temporal compromise between the gas-dominant and particle-dominant mechanisms results in the formation of mesoscale heterogeneous structure and its dynamic evolution, and therefore, the existence of a stability condition at mesoscale (Region D), i.e. $N_{st}=min$.

The two dominant mechanisms in gas-solid flows result in two possible physical states at microscale, each of which has similar dynamics. Therefore, the two different physical states should be defined as the two interpenetrating continua in mesoscience-based structural theory for gas-solid flows, instead of defining the gas and particles as the two interpenetrating continua in two-fluid model, a brief summary of which is reported in Appendix A. After deciding how to define the fluids in the mesoscience-based structural theory, it is necessary to determine the microscale governing equations (local instantaneous Navier-stokes equation for single phase flow is the microscale governing equations in two-fluid model), so that they can be averaged to formulate the mesoscience-based structural theory.

\subsection{Microscale governing equations}
From the analysis of the behavior of stability criterion of gas-solid flow at different scales and the proof of the existence of stability condition in gas-solid flow shown in Figure \ref{FigproofEMMS}, it is clear that although the characteristic spatiotemporal scale of a microscale point remains elusive \citep{ge2011meso}, it is definitely larger than a microscale point in two-fluid model, since the physical state at a point, for example Point A in Figure \ref{FigproofEMMS}, is already a gas-solid mixture \citep{li2004multi,zhang2005simulation}. On the other hand, a point should be significantly smaller than the characteristic spatiotemporal scale of mesoscale structures, such as bubbles and particle clusters. In view of the fact that the characteristic length scale of mesoscale particle clustering structures spans from few particle diameters to reactor diameter \citep{zou1994cluster,li1995cluster}, researchers might be wondering if such a characteristic spatiotemporal scale for a point exists. Fortunately, the averaging process is free from the requirement of scale separation since ensemble averaging can be used, which does not need the assumption of scale separation \citep{drew1983mathematical,prosperetti1998ensemble}.
Therefore, the first issue is to derive the microscale governing equations for the fluid consisted of gas-solid mixture within the framework of mesoscience-based structural theory. Each fluid, which is the physical realization of one dominant mechanism, is assumed to be described as a continuum that is governed by partial differential equations of continuum mechanics. The dilute-phase fluid and the dense-phase fluid are separated by mesoscale interfaces, such as the interface between bubbles and emulsion phase in bubbling fluidized beds.
The microscale governing equations of the dilute-phase fluid and the dense-phase fluid can then be further averaged to derive the mesoscience-based structural theory.

In order to derive the microscale governing equations for gas-solid mixture, the averaging method of standard two-fluid model is followed. Since the derivation of standard two-fluid model from the local instantaneous equations for gas and particles has been well documented \citep{anderson1967fluid,drew1983mathematical,gidaspow1994multiphase,prosperetti1998ensemble,jackson2000dynamics,drew2006theory}, the resultant two-fluid model, which is an averaging result of local instantaneous Navier-stokes equation for single phase flow and is the equations (\ref{a1}) to (\ref{a14}) in the Appendix A, are directly used here.
From two-fluid model, the microscale governing equations for gas-solid mixture can be derived. By adding the continuum equations for gas phase and particle phase within the dilute phase, it is easy to obtain the mixture equation of mass conservation for the dilute phase
\begin{equation}
\begin{split}
&\frac{\partial \rho_f}{\partial t}
+\nabla\cdot(\rho_f{\bf{u}}_f)=0,
\end{split}
\label{B1}
\end{equation}
where the density and velocity of dilute phase $\rho_f$ and ${\bf{u}}_f$ are respectively defined as follow
\begin{equation}
\begin{split}
\rho_f=\rho_g\varepsilon_{gf}+\rho_p(1-\varepsilon_{gf}),
\end{split}
\label{a146}
\end{equation}
\begin{equation}
\begin{split}
{\bf{u}}_f=\frac{{\rho_g\varepsilon_{gf}{\bf{u}}_{gf}+\rho_p(1-\varepsilon_{gf}){\bf{u}}_{pf}}}{\rho_g\varepsilon_{gf}+\rho_p(1-\varepsilon_{gf})},
\end{split}
\label{B3}
\end{equation}
where $\rho_g$ is the density of gases, $\varepsilon_{gf}$ is the voidage inside the dilute phase, ${\bf{u}}_{gf}$ is the gas velocity inside the dilute phase, ${\bf{u}}_{pf}$ is the particle velocity inside the dilute phase.
Similarly, the mixture equation of mass conservation for the dense phase is
\begin{equation}
\begin{split}
&\frac{\partial \rho_d}{\partial t}
+\nabla\cdot(\rho_d{\bf{u}}_d)=0,
\end{split}
\label{B2}
\end{equation}
where $\rho_d$ and ${\bf{u}}_d$ are respectively the density and velocity of the dense phase, which are defined as follow
\begin{equation}
\begin{split}
\rho_d=\rho_g\varepsilon_{gd}+\rho_p(1-\varepsilon_{gd}),
\end{split}
\label{a147}
\end{equation}
\begin{equation}
\begin{split}
{\bf{u}}_d=\frac{{\rho_g\varepsilon_{gd}{\bf{u}}_{gd}+\rho_p(1-\varepsilon_{gd}){\bf{u}}_{pd}}}{\rho_g\varepsilon_{gd}+\rho_p(1-\varepsilon_{gd})},
\end{split}
\label{B4}
\end{equation}
where $\varepsilon_{gd}$ is the voidage inside the dense phase, ${\bf{u}}_{gd}$  and ${\bf{u}}_{pd}$ are respectively the gas and particle velocities inside the dense phase. $\varepsilon_{gf}=1-\varepsilon_{pf}$ and $\varepsilon_{gd}=1-\varepsilon_{pd}$ need to be determined by some means, where $\varepsilon_{pf}$ and $\varepsilon_{pd}$ are the solid volume fractions inside the dilute phase and the dense phase, respectively. Note that the notions with subscripts $g$ and $p$ mean the variables of gas and particles, and the notions with subscripts $f$ and $d$ mean the variables of dilute phase and dense phase.

The mixture momentum equations can be obtained in a same way. The equations (\ref{a13}) and (\ref{a14}) are firstly rewritten in a different form without any changes of their physics \citep{wang2020continuum}
\begin{equation}
\begin{split}
\frac{\partial(\varepsilon_g\rho_g{\bf{u}}_g)}
{\partial t}
+\nabla\cdot(\varepsilon_g\rho_g{\bf{u}}_g{\bf{u}}_g)
=-\nabla (\varepsilon_g p_g)+\nabla\cdot(\varepsilon_g{\bm{\tau}}_g)
+\varepsilon_g\rho_g{\bf{g}}-{\bf{M}}_{i}
\end{split}
\label{B5}
\end{equation}
and
\begin{equation}
\begin{split}
\frac{\partial(\varepsilon_p\rho_p{\bf{u}}_p)}
{\partial t}
+\nabla\cdot(\varepsilon_p\rho_p{\bf{u}}_p{\bf{u}}_p)
=-\nabla (\varepsilon_p p_g)-\nabla p_p+\nabla\cdot(\varepsilon_p{\bm{\tau}}_p)
+\varepsilon_p\rho_p{\bf{g}}+{\bf{M}}_{i},
\end{split}
\label{B6}
\end{equation}
where $p_p$ is the so-called granular pressure, ${\bm{\tau}}_p$ is the viscous stress tensor of particles, ${\bf{M}}_{i}$ represents the interfacial momentum transfer rate, which is different to the drag force ${\bf{F}}_{drag}$. For a detailed discussion of the equality between equations (\ref{a13}) and (\ref{a14}) and equations (\ref{B5}) and (\ref{B6}), see the recent review of \citet{wang2020continuum}.
Via the addition of the momentum conservation equations for gas phase and particle phase within the dilute phase, the mixture momentum equation for the dilute phase is
\begin{equation}
\begin{split}
\frac{\partial(\rho_f{\bf{u}}_f)}
{\partial t}
+\nabla\cdot(\rho_f{\bf{u}}_f {\bf{u}}_f)
=-\nabla p_f +\nabla\cdot {\bm{\tau}}_f
+\rho_f{\bf{g}},
\end{split}
\label{B7}
\end{equation}
where the pressure of dilute phase is
\begin{equation}
\begin{split}
p_f=p_g+p_{pf},
\end{split}
\label{B7-7}
\end{equation}
and the effective shear stress tensor of dilute phase is \citep{gidaspow1994multiphase,manninen1996mixture,wang2012emms}
\begin{equation}
\begin{split}
{\bm{\tau}}_f=\varepsilon_{gf}{\bm{\tau}}_{gf}+\varepsilon_{pf}{\bm{\tau}}_{pf}-[\varepsilon_{gf}\rho_g {\bf{u}}_{gf,rel}{\bf{u}}_{gf,rel}+\varepsilon_{pf}\rho_p {\bf{u}}_{pf,rel}{\bf{u}}_{pf,rel}],
\end{split}
\label{B8}
\end{equation}
where the relative velocity with respect to the mixture velocity is defined as ${\bf{u}}_{i,rel}={\bf{u}}_{i}-{\bf{u}}_{f}$ with $i=gf$ or $i=pf$. The effective viscous stress tensor of dilute phase can be closed following the mixture theory of multicomponent flow \citep{curtiss1996multicomponent,Bird2002transport,datta2010continuum}
\begin{equation}
\begin{split}
{\bm{\tau}}_f=\mu_f(\nabla {\bf{u}}_{f}+ \nabla {\bf{u}}_{f}^T)-\bigg(\frac{2}{3}\mu_f -\lambda_f \bigg)\nabla \cdot {\bf{u}}_{f} {\bf{I}},
\end{split}
\label{B9}
\end{equation}
where $\mu_f$ and $\lambda_f$ are respectively the effective shear and bulk viscosities of dilute phase, $T$ denotes the transpose, ${\bf{I}}$ is the unit tensor. Note that the definition of ${\bm{\tau}}_f$ as presented in equations (\ref{B8}) and (\ref{B9}) is debatable, for example, \citet{manninen1996mixture} and the mixture model available in FLUENT$^\circledR$ have singled the diffusive stress tensor (the last term in equation (\ref{B8})) out and developed a model for  ${\bf{u}}_{i,rel}$ to close it. In addition to the simplicity, present study will use an empirical correlation to close the effective viscosity of whole suspension, therefore, the effect of the diffusive stress tensor should already be included, since it is also part of the particle contribution to the effective suspension stress \citep{nott2011suspension}. Moreover, from a practical viewpoint the contribution of the diffusive stress tensor is usually negligible \citep{jamshidi2021roles}, when it is compared to the contribution of the phasic effective stress tensors. Similarly, the mixture momentum equation for the dense phase is
\begin{equation}
\begin{split}
\frac{\partial(\rho_d{\bf{u}}_d)}
{\partial t}
+\nabla\cdot(\rho_d{\bf{u}}_d {\bf{u}}_d)
=-\nabla p_d+\nabla\cdot {\bm{\tau}}_d
+\rho_d{\bf{g}},
\end{split}
\label{B10}
\end{equation}
with
\begin{equation}
\begin{split}
p_d=p_g+p_{pd},
\end{split}
\label{B10-1}
\end{equation}
and
\begin{equation}
\begin{split}
{\bm{\tau}}_d=\varepsilon_{gd}{\bm{\tau}}_{gd}+\varepsilon_{pd}{\bm{\tau}}_{pd}-[\varepsilon_{gd}\rho_g {\bf{u}}_{gd,rel}{\bf{u}}_{gd,rel}+\varepsilon_{pd}\rho_p {\bf{u}}_{pd,rel}{\bf{u}}_{pd,rel}],
\end{split}
\label{B11}
\end{equation}
where ${\bf{u}}_{i,rel}={\bf{u}}_{i}-{\bf{u}}_{d}$ with $i=gd$ or $i=pd$. Furthermore, the constitutive relationship for ${\bm{\tau}}_d$ is similar to equation (\ref{B9}).

Experimental studies have found that the surface tension force of noncohesive granular flow, which results from the granular pressure difference at the interface \citep{clewett2012emergent}, is essentially zero \citep{cheng2007collective,cheng2008towards} or ultralow \citep{amarouchene2008capillarylike,prado2011experimental,clewett2012emergent,luu2013capillarylike}. More importantly, there are direct evidences of the equality of granular pressures of dilute phase $p_{pf}$ and dense phase $p_{pd}$ \citep{clewett2016minimization,clewett2019reduced,lu2022phase,lu2022compressibility}, therefore, it seems reasonable to assume
\begin{equation}
p_f=p_d.
\label{C17}
\end{equation}
It should be addressed here that considering the facts that the so-called equal pressure assumption \citep{stewart1984two} has been used in the derivation of two-fluid model presented in the Appendix A and $p_{pf}=p_{pd}$ is not rigorously valid as just discussed, equation (\ref{C17}) is better being recognized as a basic assumption.

The equations (\ref{B1}), (\ref{B2}), (\ref{B7}) and (\ref{B10}) are the microscale governing equations for gas-solid mixtures, which can be further averaged to formulate the mesoscience-based structural theory. It should be emphasized here that (i) the microscale interfaces between gas and particles at the particle surfaces have been averaged out, but the mesoscale interfaces between dilute phase and dense phase are waiting for being averaged; (ii) in the standard mixture model for fluid-particle suspensions \citep{manninen1996mixture,jamshidi2019closure}, another mass balance equation for each fluid is required to track the solid volume fractions ($\varepsilon_{pf}$ and $\varepsilon_{pd}$), and therefore, the densities of dilute and dense phases ($\rho_f$ and $\rho_d$); and (iii) the microscale governing equations of gas-solid mixtures can also derived following the extensive studies on the continuum mixture theories of fluids and solids \citep{bowen1980incompressible,bowen1982compressible,baer1986two,rajagopal1995mechanics,liu2014solid}.

\subsection{Formulation of mesoscience-based structural theory}
\subsubsection{Ensemble-averaged conservation equations}
The analysis of the behavior of stability condition in Figure \ref{FigproofEMMS} has shown that one dominant mechanism can fully fulfil its extremum tendency locally and instantaneously, and the gas-dominant and particle-dominant mechanisms can only realize its extremum tendency alternatively at a spatiotemporal point $({\bf{x}},t)$, where ${\bf{x}}$ is position and $t$ is time. Therefore, it is reasonable to introduce the dominant mechanism indicator function $X_k({\bf{x}},t)$, which is defined as follows:
\begin{equation}
\begin{split}
X_k({\bf{x}},t)=
    \begin{cases}
        1 & \text{if particle-dominant mechanism is realized at} \  ({\bf{x}},t)  \\
        0 & \text{if gas-dominant mechanism is realized at} \  ({\bf{x}},t).
      \end{cases}
\end{split}
\label{C1}
\end{equation}

Let us define $h({\bf{x}},t)$ as an exact microscale field, then $\bar{h}({\bf{x}},t)$ is the corresponding averaged field. From equation (\ref{C1}), it is easy to have
\begin{equation}
f_k=\bar{X}_k,
\label{C1-1}
\end{equation}
where $f_k$ is the volume fraction of phase $k$ which is also the fraction of the realization of particle- or gas-dominant mechanism, and $k$ can be $f$ or $d$, which correspondingly means dilute phase or dense phase. Clearly, if the system is fully dominated by the particle-dominant mechanism, then $f_d=1$, which means the gas-solid system is operated at the fixed bed regime.  On the other hand, if gas-dominant mechanism is overwhelming, then $f_f=0$, which means the system is operated at the dilute transportation regime. Furthermore, the volume fractions should satisfy the geometrical constraint, that is,
\begin{equation}
\begin{split}
f_f+f_d=1.
\end{split}
\label{C1-2}
\end{equation}
The averaging process is assumed to satisfy the Reynolds's rule, the Leibniz's rule and the Gauss's rule \citep{drew1983mathematical,enwald1996eulerian}:
\begin{equation}
\overline{h+g}=\bar{h}+\bar{g},
\label{C2}
\end{equation}
\begin{equation}
\overline{\bar{h}g}=\bar{h}\bar{g},
\label{C3}
\end{equation}
\begin{equation}
\overline{\text{constant}}=\text{constant},
\label{C4}
\end{equation}
\begin{equation}
\overline{\frac{\partial h}{\partial t}}=\frac{\partial \bar{h}}{\partial t},
\label{C5}
\end{equation}
\begin{equation}
\overline{\nabla h}=\nabla \bar{h},
\label{C6}
\end{equation}
\begin{equation}
\overline{\nabla \cdot h}=\nabla \cdot \bar{h}.
\label{C7}
\end{equation}
Finally, the material derivative of the dominant mechanism indicator following the interface is a critical fundamental relation in the averaging process \citep{drew1983mathematical,drew1993analytical}:
\begin{equation}
\begin{split}
\frac{\partial X_k}{\partial t}+{\bf{u}}_i \cdot \nabla X_k=0,
\end{split}
\label{C8}
\end{equation}
where ${\bf{u}}_i$ is the velocity of mesoscale interface. This equation is easy to be understood: If a point is not on the interface, or physically, it is dominated by one dominant mechanism, then either $X_k=1$ or $X_k=0$. In either case, the partial derivatives of $X_k$ with respect to time and space are both zero, therefore, equation (\ref{C8}) is valid. If a point is on the interface and move with it, $X_k$ is a constant jump, then its material derivative is zero.

Multiplying equation (\ref{B1}) by the dominant mechanism indicator function $X_k$ and then averaging using the mathematical techniques developed by Drew \citep{drew1983mathematical,enwald1996eulerian,drew2006theory}, we have the averaged mass equation of dilute phase and dense phase:
\begin{equation}
\frac{\partial (f_k \tilde{\rho}_k)}{\partial t}+ \nabla \cdot (f_k \tilde{\rho}_k \hat{{\bf{u}}}_k)
=\overline{{\rho}_k({\bf{u}}_k-{\bf{u}}_{i,k}) \cdot \nabla X_k},
\label{C9}
\end{equation}
\begin{comment}
\begin{equation}
\frac{\partial f_f \tilde{\rho}_f}{\partial t}+ \nabla \cdot (f_f \tilde{\rho}_f \hat{{\bf{u}}}_f)
=\overline{{\rho}_f({\bf{u}}_f-{\bf{u}}_{i,f}) \cdot \nabla X_f},
\label{C9}
\end{equation}
similarly, the averaged mass equation of dense phase can be derived from equation (\ref{B2}):
\begin{equation}
\frac{\partial f_d \tilde{\rho}_d}{\partial t}+ \nabla \cdot (f_d \tilde{\rho}_d \hat{{\bf{u}}}_d)
=\overline{{\rho}_d({\bf{u}}_d-{\bf{u}}_{i,d}) \cdot \nabla X_d},
\label{C10}
\end{equation}
\end{comment}
where the $X_k$-weighted averaged density of phase $k$ is defined as $\tilde{\rho}_k=\overline{X_k \rho_k}/f_k$, and the mass-weighted or Favre averaged velocity of phase $k$ is defined as $\hat{{\bf{u}}}_k=\overline{X_k \rho_k {\bf{u}}_k}/(f_k \tilde{\rho}_k)$.

Multiplying equations  (\ref{B7}) and (\ref{B10}) by the dominant mechanism indicator function $X_k$ and then averaging, we have the averaged momentum conservation equation of dilute phase and dense phase:
\begin{equation}
\frac{\partial (f_k \tilde{\rho}_k \hat{{\bf{u}}}_k)}{\partial t}+ \nabla \cdot (f_k \tilde{\rho}_k \hat{{\bf{u}}}_k \hat{{\bf{u}}}_k)
= -\nabla (f_k \tilde{p}_k) + \nabla\cdot [f_k (\tilde{\bm{\tau}}_k + \tilde{\bm{\tau}}_k^{\text{Re}})]
 + f_k \tilde{\rho}_k{\bf{g}} + \overline{{{[\rho}_k\bf{u}}_k({\bf{u}}_k-{\bf{u}}_{i,k})-{\bm{\tau}}_k ] \cdot \nabla X_k}
  + \overline{p_k  \nabla X_k},
\label{C12}
\end{equation}
\begin{comment}
\begin{equation}
\frac{\partial f_f \tilde{\rho}_f \hat{{\bf{u}}}_f}{\partial t}+ \nabla \cdot (f_f \tilde{\rho}_f \hat{{\bf{u}}}_f \hat{{\bf{u}}}_f)
= -\nabla (f_f \tilde{p}_f) -\nabla (f_f \tilde{p}_{pf}) + \nabla\cdot [f_f (\tilde{\bm{\tau}}_f + \tilde{\bm{\tau}}_f^{\text{Re}})]
 + f_f \tilde{\rho}_f{\bf{g}} + \overline{{{[\rho}_f\bf{u}}_f({\bf{u}}_f-{\bf{u}}_{i,f})-{\bm{\tau}}_f ] \cdot \nabla X_f}
  + \overline{(p_f + p_{pf}) \nabla X_f},
\label{C12}
\end{equation}
\begin{equation}
\frac{\partial f_d \tilde{\rho}_d \hat{{\bf{u}}}_d}{\partial t}+ \nabla \cdot (f_d \tilde{\rho}_d \hat{{\bf{u}}}_d \hat{{\bf{u}}}_d)
= -\nabla (f_d \tilde{p}_d) -\nabla (f_d \tilde{p}_{pd}) + \nabla\cdot [f_d (\tilde{\bm{\tau}}_d + \tilde{\bm{\tau}}_d^{\text{Re}})]
 + f_d \tilde{\rho}_d{\bf{g}} + \overline{{[{\rho}_d\bf{u}}_d({\bf{u}}_d-{\bf{u}}_{i,d}) -{\bm{\tau}}_d ] \cdot \nabla X_d}
 + \overline{(p_d + p_{pd}) \nabla X_d},
\label{C12}
\end{equation}
\end{comment}
where the Reynolds (or sub-grid-scale) stress tensors are defined as
\begin{equation}
\tilde{\bm{\tau}}_k^{\text{Re}}= \tilde{\rho}_k \hat{{\bf{u}}}_k \hat{{\bf{u}}}_k-\frac{\overline{X_k \rho_k {{\bf{u}}}_k{{\bf{u}}}_k}}{{f}_k}.
\label{C13}
\end{equation}
and the laminar shear stress tensor is
\begin{equation}
\begin{split}
\tilde{\bm{\tau}}_k=\mu_k(\nabla \hat{{\bf{u}}}_{k}+ \nabla \hat{{\bf{u}}}_{k}^T)-\bigg(\frac{2}{3}\mu_k -\lambda_k \bigg)\nabla \cdot \hat{{\bf{u}}}_{k} {\bf{I}}.
\end{split}
\label{C14}
\end{equation}
In present study, we assume that (i) the interface between dilute phase and dense phase is sharp, which means that $\nabla X_k$ is zero everywhere, except on the interface and $\nabla X_k={\bf{n}}_k\frac{\partial X}{\partial n}$, where ${\bf{n}}_k$ is the unit normal exterior to the fluid $k$ with ${\bf{n}}_f=-{\bf{n}}_d$ and $\frac{\partial X}{\partial n}=\delta ({\bf{x}}-{\bf{x}}_i,t)$ is the Dirac's delta function; (ii) there is no storage or accumulation of mass and momentum at an interface; and (iii) the surface tension force is zero, as has been discussed. Therefore, the averaged interfacial mass and momentum constraints are \citep{ishii2010thermo}
\begin{equation}
\sum_{k=1}^2 \Gamma_k= \sum_{k=1}^2  \overline{{\rho}_k({\bf{u}}_k-{\bf{u}}_{i,k}) \cdot \nabla X_k}=0,
\label{C14}
\end{equation}
\begin{equation}
\sum_{k=1}^2 {\bf{M}}_k=\sum_{k=1}^2 \overline{{{[\rho}_k\bf{u}}_k({\bf{u}}_k-{\bf{u}}_{i,k})-{\bm{\tau}}_k ] \cdot \nabla X_k}
  + \overline{p_k \nabla X_k}=0,
\label{C15}
\end{equation}
where $\Gamma_k$ is the defined interfacial mass transfer rate, and ${\bf{M}}_k$ is the interfacial momentum transfer rate.

Following previous studies \citep{drew1983mathematical,enwald1996eulerian,jackson2000dynamics,drew2006theory,ishii2010thermo}, it is customary to define
\begin{equation}
\Gamma_k {\bf{u}}_{i,k}=\overline{{\rho}_k{\bf{u}}_k({\bf{u}}_k-{\bf{u}}_{i,k}) \cdot \nabla X_k}
\label{C16}
\end{equation}
as the interfacial momentum transfer caused by interfacial mass transfer, where ${\bf{u}}_{i,k}$ is the interfacial velocity of the $k$th fluid.
Furthermore, the term related to the interfacially averaged stress tensor, which includes the interfacially averaged pressure $\bar{p}_{i,k}$ and the interfacially averaged shear stress $\bar{\bm{\tau}}_{i,k}$, is usually singled out from the interfacial momentum transfer, therefore, we have
\begin{equation}
{\bf{M}}_k=\Gamma_k {\bf{u}}_{i,k} + \bar{p}_{i,k} \nabla f_k -\bar{\bm{\tau}}_{i,k} \cdot \nabla f_k + {\bf{M}}_k^{F},
\label{C18}
\end{equation}
with
\begin{equation}
{\bf{M}}_k^{F}=\overline{(p_k-\bar{p}_{i,k})\nabla X_k-{({\bm{\tau}}_k-\bar{\bm{\tau}}_{i,k})  \cdot \nabla X_k}},
\label{C19}
\end{equation}
where ${\bf{M}}_k^{F}$ is referred as the interfacial force density \citep{anderson1967fluid,drew1993analytical}, which includes various forces that are usually included in numerical simulations, such as drag force and added mass force. Furthermore, $\bar{\bm{\tau}}_k \cdot \nabla f_k$ can be neglected for disperse flows \citep{ishii1984two}. It is also customary to separate $-\nabla (f_k \tilde{p}_k)$ as $-f_k\nabla (\tilde{p}_k)$ and $-\tilde{p}_k \nabla f_k$, and then combine $-\tilde{p}_k \nabla f_k$ and $\bar{p}_{i,k} \nabla f_k$ as $(\bar{p}_{i,k}-\tilde{p}_k) \nabla f_k$, which is also negligible for the gas-solid flows studied here \citep{drew1983mathematical,ishii2010thermo}, since the surface tension force is essentially zero.

By dropping off the bar, hat and tilde for the simplicity of notation and implementing the discussions of interfacial mass and momentum transfer rates, we have the final governing equations of the mesoscience-based structural theory. The mass conservation equations of dilute phase and dense phase are
\begin{equation}
\begin{split}
&\frac{\partial (f_f\rho_f)}{\partial t}
+\nabla\cdot(f_f\rho_f{\bf{u}}_f)=\Gamma_f
\end{split}
\label{a11}
\end{equation}
and
\begin{equation}
\begin{split}
&\frac{\partial (f_d\rho_d)}{\partial t}
+\nabla\cdot(f_d\rho_d{\bf{u}}_d)=\Gamma_d.
\end{split}
\label{a22}
\end{equation}
The momentum conservation equations are
\begin{equation}
\begin{split}
\frac{\partial(f_f\rho_f{\bf{u}}_f)}
{\partial t}
+\nabla\cdot(f_f\rho_f{\bf{u}}_f{\bf{u}}_f)
=-f_f\nabla p_f+\nabla\cdot(f_f{\bm{\tau}}_f)
+f_f\rho_f{\bf{g}}+{\bf{M}}_{f}^F+\Gamma_f {\bf{u}}_{i,f}
\end{split}
\label{a133}
\end{equation}
and
\begin{equation}
\begin{split}
\frac{\partial(f_d\rho_d{\bf{u}}_d)}
{\partial t}
+\nabla\cdot(f_d\rho_d{\bf{u}}_d{\bf{u}}_d)
=-f_d\nabla p_d+\nabla\cdot(f_d{\bm{\tau}}_d)
+f_d\rho_d{\bf{g}}+{\bf{M}}_{d}^F+\Gamma_d {\bf{u}}_{i,d}.
\end{split}
\label{a144}
\end{equation}

It can be seen that there are twelve state variables $f_f$, $f_d$, $\rho_f$, $\rho_d$, $p_f$, $p_d$, ${\bf{u}}_f$ and ${\bf{u}}_d$, but only ten conservation equations can be formulated (equations (\ref{C17}), (\ref{C1-2}), and (\ref{a11})-(\ref{a144})). Therefore, {the conservation equations alone are insufficient to uniquely specify the spatiotemporal dynamics of heterogeneous gas-solid flows, even the constitutive relationships can be provided}, including the interphase mass transfer rates between phase $\Gamma_f$ and $\Gamma_d$, the effective stress tensors of dilute phase ${\bm{\tau}}_f$ and dense phase ${\bm{\tau}}_d$, the interfacial force densities ${\bf{M}}_{f}^F$ and ${\bf{M}}_{d}^F$, and the momentum transfer rates due to the interphase mass transfer $\Gamma_f {\bf{u}}_{i,f}$ and $\Gamma_d {\bf{u}}_{i,d}$. The lack of sufficient numbers of conservation equations originates from the fact that although the densities of gas and particles, $\rho_g$ and $\rho_p$, are known, the densities of dilute phase and dense phase, $\rho_f$ and $\rho_d$, are still unknowns, because both of dilute phase and dense phase are mixtures of gas and particles with different voidages, see equations (\ref{a146}) and (\ref{a147}). Of course, it is able to formulate more equations following classic mixture model as shown in the Appendix B.

\subsubsection{Stability condition}
In addition to the conservation equations (equations (\ref{C1-2}) and (\ref{a11})-(\ref{a144})) and the basic assumption (equation (\ref{C17})), the compromise in competition between dominant mechanisms also results in a stability condition that constrains the spatiotemporal dynamics of heterogeneous gas-solid flows (for example in Region D of Figure \ref{FigproofEMMS}), as already presented in the analysis of the behavior of stability criterion at different scales and the proof of the existence of stability condition in Figure \ref{FigproofEMMS}. Since the conservation equations have averaged over the mesoscale interfaces,
the new physics at mesoscale that is the existence of stability condition should be an indispensable part of theoretical formulation, and is actually the core of mesoscience-based structural theory.

The original stability condition $N_{st}=min$ is not a function of space and time, present study necessitates the proper extension of $N_{st}=\frac{W_{st}}{\varepsilon_s\rho_p}=min$. Following the idea of axial and/or radial EMMS model \citep{li1990application,li1994particle,hu2013steady,hu2017quantifying}, we define the stability condition as
\begin{equation}
\begin{split}
\bar{N}_{st}(t)=\frac{\int \varepsilon_s({\bf{x}},t){N}_{st}({\bf{x}},t)d{\bf{x}}}{\int \varepsilon_s({\bf{x}},t)d{\bf{x}}}
=\frac{\int {W}_{st}({\bf{x}},t)d{\bf{x}}}{\rho_p \int\varepsilon_s({\bf{x}},t)d{\bf{x}}}=min,
\end{split}
\label{a145}
\end{equation}
where $\varepsilon_s=f_f\varepsilon_{pf}+f_d\varepsilon_{pd}$ is the solid concentration. Clearly, $\bar{N}_{st}(t)$, which is the mass-specific density of $W_{st}$ at time $t$, is an intensive variable.
It is interesting to note that (i) the density of entropy production, which is also an intensive variable, is integrated over the whole system to prove the minimum entropy production principle \citep{prigogine1967introduction,de2013non}. Therefore, the entropy production in classic nonequilibrium thermodynamics is an extensive and additive variable, which is qualitatively different with the intensive variable $\bar{N}_{st}(t)$;
and (ii) in the numerical solution of mesoscience-based structural theory, it might be possible to use a local stability condition $N_{st}({\bf{x}},t)=\frac{W_{st}({\bf{x}},t)}{\rho_p\varepsilon_s({\bf{x}},t)}=min$, the computational cost should then be extremely high.

\begin{figure}[!htb]
\centering
\includegraphics[width=0.8\textwidth]{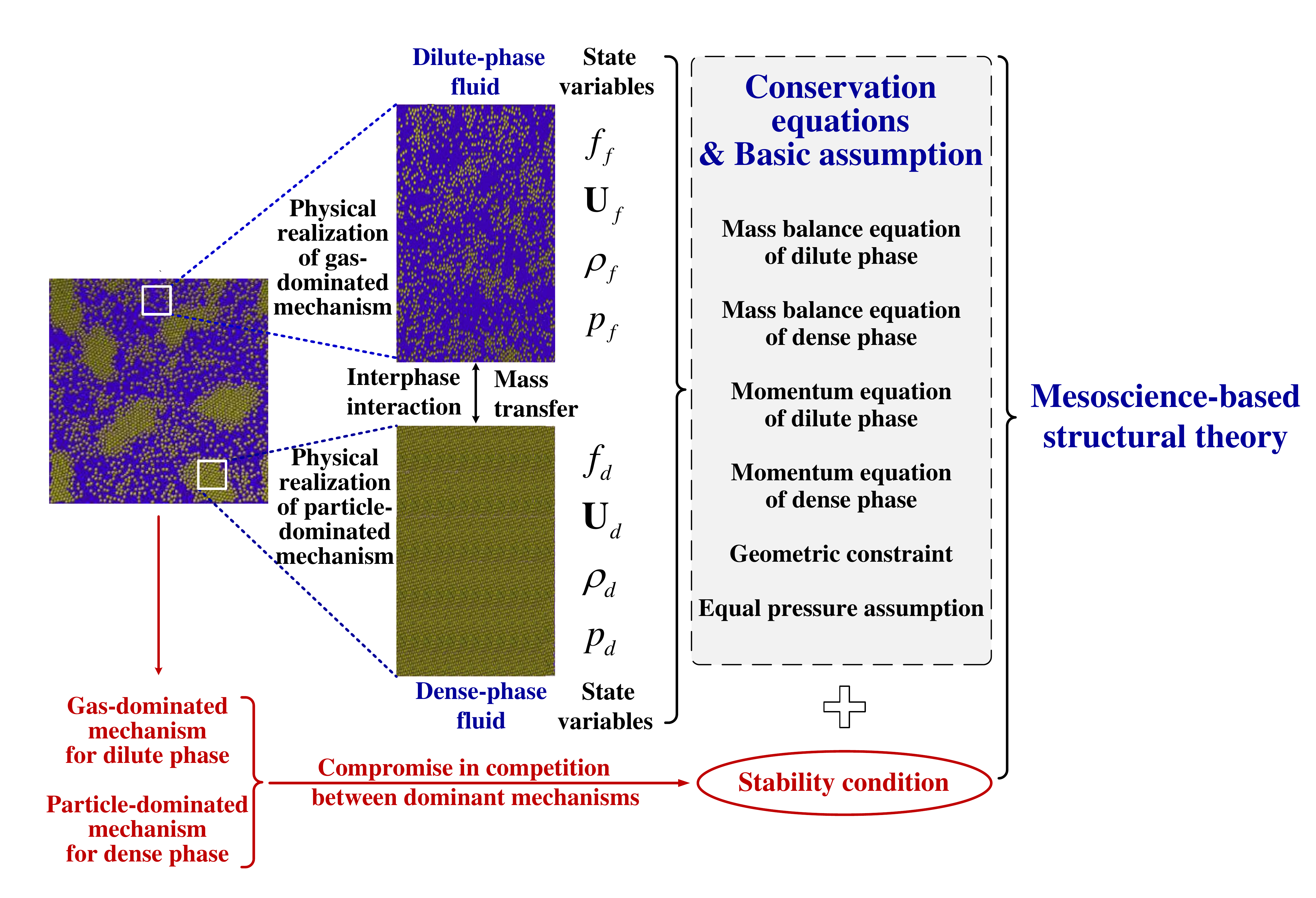}
\caption{Framework of mesoscience-based structural theory for heterogeneous gas-solid flows}
\label{figFramework}
\end{figure}

As summarized in Figure \ref{figFramework}, mesoscience-based structural theory  defines the two interpenetrating fluids according to the gas-dominant and particle-dominant mechanisms existed in complex gas-solid systems, conservation equations and a basic assumption are then formulated (equations (\ref{C17}), (\ref{C1-2}), and (\ref{a11})-(\ref{a144})); At the same time, a stability condition is also formulated from the compromise in competition between dominant mechanisms (equation (\ref{a145})). {The conservation equations, the basic assumption and the stability condition together form the mesoscience-based structural theory . Therefore, the mesoscience-based structural theory  is mathematically formulated as partial differential equations constrained dynamic optimization}.

\subsection{Constitutive relationship}
In the mesoscience-based structural theory, constitutive relationships for the effective shear viscosities of dilute phase and dense phase $\mu_k$, the interphase mass transfer rate $\Gamma_k$ and the interfacial force density ${\bf{M}}_{k}^F$ are needed. In the classical two-fluid model, the bulk viscosity is routinely predicted by the kinetic theories of granular flow \citep{gidaspow1994multiphase}. In present study, empirical correlations are used to close the transport coefficients in the particulate phase stress, in this case the bulk viscosity $\lambda_k$ is usually neglected in state-of-the-art studies \citep{wang2020continuum}. However, very few studies are available if needed \citep{brady2006bulk,swaroop2007bulk}. As will be shown, all the constitutive laws take the advantages that the interactions within the dense phase and the dilute phase, as well as the interaction between them are all closed using models and correlations assuming a homogeneous structure, leaving the proper prediction of characteristic length scale of mesoscale structures (cluster diameter and/or bubble size) as the major challenge. On the other hand, the mesoscience-based structural theory  necessities a model for the interphase mass transfer even only the hydrodynamics is concerned, which is a disadvantage as compared to two-fluid model.

\subsubsection{Effective shear viscosity}
The particles interact mainly via hydrodynamic interactions at low solid volume fraction, the fluid mechanics of such dilute suspensions at the low Reynolds number limit are well understood, primarily due to the seminal works of Einstein \citep{einstein1905neue,einstein1911berichtigung} and Batchelor \citep{batchelor1970stress,batchelor1972determination}, which results in the effective viscosity of suspension is $\mu_k=\mu_g(1+2.5\varepsilon_{pk}+5\varepsilon_{pk}^2)$. Particle-particle collisions and sustained contacts become critical in dense or concentrated fluid-particle suspensions, the rheology of which is much more difficult to be understood and remains a hot topic of researches after extensive studies \citep{bagnold1954experiments,krieger1959mechanism,thomas1965transport,krieger1972rheology,jeffrey1976rheological,stickel2005fluid,
mewis2012colloidal,brown2014shear,denn2014rheology,guazzelli2018rheology,ness2022physics}.
On the other hand, the effective viscosity of fluidized beds has been extensively measured \citep{kramers1951viscosity,grace1970viscosity,schugerl1971rheological,bakhtiyarov1998fluidized,gibilaro2007apparent,colafigli2009apparent,
amin2021fluidization,amin2022fall}.
Present study is not going to systematically summarize the state-of-the-art, but reporting representative correlations available in literature, which have been or can be used in closing the constitutive relationships of mesoscience-based structural theory .

For systems consisting of non-Brownian, rigid, neutrally buoyant spheres suspended in Newtonian fluids, the effective shear viscosity can be expressed as \citep{krieger1959mechanism,krieger1972rheology}
\begin{equation}
\begin{split}
\mu_k=\mu_g\bigg(1-\frac{\varepsilon_{pk}}{\varepsilon_{pc}}\bigg)^{-\frac{5}{2}\varepsilon_{pc}},
\end{split}
\label{d1}
\end{equation}
where $\varepsilon_{pc}$ is the solid volume fraction at the jamming transition point, which is in general different to the solid volume fraction at the random packing state ($\varepsilon_{s,max}\approx0.636$). The value of $\varepsilon_{pc}$, which is critical for the correct prediction of suspension viscosity but is difficult to be determined, mainly depends on the particle size distribution and particle surface interactions \citep{guazzelli2018rheology}. Therefore, $\varepsilon_{pc}$ is usually used as an adjustable parameter \citep{stickel2005fluid}, although efforts have also been made to develop shear-dependent $\varepsilon_{pc}$ \citep{wildemuth1984viscosity,zhou1995yield}, but then many more empirical constants are needed. Clearly, equation (\ref{d1}) means that the fluid-particle suspension can be seen as a Newtonian fluid with an effective viscosity increasing with increasing solid concentration and diverging at $\varepsilon_{pc}$. This kind of correlations are best used in the Stokes flow limit where the particle and fluid inertias are essentially zero, or used in the so-called viscous regime. Although there are many other correlations available in literature with a clearer physical insight \citep{boyer2011unifying}, this classic correlation offers a satisfactory prediction of the effective viscosity of fluid-particle suspensions \citep{stickel2005fluid,guazzelli2018rheology} and has the advantages of succinct expression and that it is valid for whole range of solid volume fraction.

When the particle inertia is important but the inertia of gas can still be neglected, an empirical correlation was proposed recently on the basis of extensive fluidization and sedimentation experiments \citep{amin2021fluidization}:
\begin{equation}
\begin{split}
\mu_k=\frac{\mu_g}{(e^{-3}+0.08)}\bigg(2\frac{\text{St}/45}{1+\text{St}/45}+1\bigg)\bigg[e^{-3\big(1- \frac{\varepsilon_{pk}}{\varepsilon_{pc}}\big)} + 0.08\bigg(1- \frac{\varepsilon_{pk}}{\varepsilon_{pc}}\bigg)^{-\frac{2}{3}}\bigg],
\end{split}
\label{d2}
\end{equation}
where the Stokes number is defined as $\text{St}=\frac{(\rho_p+1/2\rho_g)(\rho_p-\rho_g)gd_p^3}{18\mu_g^2}$. Clearly, the effective viscosity is not only a function of $\frac{\varepsilon_{pk}}{\varepsilon_{pc}}$ as in equation (\ref{d1}) but also a function of $\text{St}$ which reflects the effects of particle inertia. On the other hand, for a system with given gas and particle properties, it is still only a function of $\frac{\varepsilon_{pk}}{\varepsilon_{pc}}$, since $\text{St}$ is a constant.

Macroscopic shear is one of the main mechanisms of gas-particle transport processes, and shear thickening at high shear rate is a common phenomena of dense suspensions \citep{bagnold1954experiments,brown2014shear,morris2020shear}. Recently, an attempt was made to develop a unified model that includes both of viscous and inertial contributions to the effective viscosity \citep{tapia2022viscous}
\begin{equation}
\begin{split}
\mu_k=\mu_g a_{\varepsilon}^2\eta_c(1+\alpha_{\varepsilon}\text{St})\bigg(1-\frac{\varepsilon_{pk}}{\varepsilon_{pc}}\bigg)^{-2}\Bigg[1+\frac{a_{\eta}}{a_{\varepsilon}}\bigg(1- \frac{\varepsilon_{pk}}{\varepsilon_{pc}}\bigg) \sqrt{\bigg(\frac{1+\alpha_{\eta}\text{St}}{1+\alpha_{\varepsilon}\text{St}}\bigg)} \Bigg]
\end{split}
\label{d3}
\end{equation}
where the empirical constants in the model are $a_{\varepsilon}=0.66$, $\eta_c=0.31$, $a_{\eta}=11.29$, $\alpha_{\eta}=0.0088$, $\alpha_{\varepsilon}=0.1$, $\varepsilon_{sc}=0.615$ and the Stokes number is now defined as $\text{St}=\frac{\rho_p d_p^2 J}{\mu_g}$, where $J$ is the shear rate. When it is implemented into three-dimensional numerical simulations, we may assume $J=\sqrt{2\bar{\bar{D}}_k : \bar{\bar{D}}_k}$ \citep{montella2021two}, where $\bar{\bar{D}}_k=\frac{1}{2}[\nabla {\bf{u}}_k + (\nabla {\bf{u}}_k)^T]$ is the strain rate tensor of phase $k$. When $\text{St}=0$ and $\varepsilon_{pk}=0$, $\mu_k\neq \mu_g$, which means the model does not have a correct dilute and Stokes limit.

Recent studies have found that the effective viscosity of fluid-particle suspension is simply the sum of viscous contribution and inertial contribution \citep{trulsson2012transition,amarsid2017viscoinertial,vo2020additive,tapia2022viscous}, then we propose an $ad  \ hoc$ engineering model here
\begin{equation}
\begin{split}
\mu_k=\mu_k^{vis} + \mu_k^{inert},
\end{split}
\label{d4}
\end{equation}
where the viscosity of viscous contribution $\mu_k^{vis}$ can be calculated using equation (\ref{d1}) and the inertial contribution $\mu_k^{inert}$ can be obtained from the extensive studies on granular flow. $\mu_k^{inert}$ itself is also the addition of kinetic and collisional contribution predicted by kinetic theory of granular flow, and the frictional contribution \citep{johnson1987frictional,forterre2008flows}:
\begin{equation}
\begin{split}
\mu_{k}^{inert}=\frac{4}{5}(1+e)\varepsilon_{pk}^2{g_0}\rho_pd_p\sqrt{\frac{\theta_{pk}}{\pi}}+\frac{5\rho_pd_p\sqrt{\theta_{pk}\pi}}{48(1+e)g_0}[1+0.8(1+e)\varepsilon_{pk}g_0]^2
+\frac{p^f\mu_i(I_s)}{\sqrt{2\bar{\bar{D}}_k : \bar{\bar{D}}_k}},
\end{split}
\label{d5}
\end{equation}
with
\begin{equation}
\begin{split}
g_0=\bigg[1-(\frac{1-\varepsilon_{gk}}{\varepsilon_{s,max}})^{\frac{1}{3}}\bigg]^{-1}, \quad  \mu_i(I_s)
=\mu_i^{st}+\frac{\mu_i^c-{\mu}_i^{st}}{I_0/I_s+1}
\quad \text{and} \quad
I_s=\frac{d_p\sqrt{2\bar{\bar{D}}_k:\bar{\bar{D}}_k}}{\sqrt{p^f/\rho_p}},
\end{split}
\label{d7}
\end{equation}
where $\mu_i^{st}$, $\mu_i^c$, $I_0$ are constants depending on material properties, and the frictional pressure $p^f$ is \citep{tapia2022viscous}
\begin{equation}
\begin{split}
p^f=\mu_g a_{\varepsilon}^2\bigg(1+\alpha_{\varepsilon}\frac{\rho_p d_p^2 \sqrt{2\bar{\bar{D}}_k:\bar{\bar{D}}_k} }{\mu_g}\bigg)\bigg(1-\frac{\varepsilon_{pk}}{\varepsilon_{pc}}\bigg)^{-2} \sqrt{2\bar{\bar{D}}_k:\bar{\bar{D}}_k}.
\end{split}
\label{d8}
\end{equation}
The granular temperature $\theta_{pk}$ in equation (\ref{d5}) can be predicted by the algebraic approximation of granular temperature equation that assumes the fluctuating energy is generated and dissipated locally \citep{van1998eulerian,musser2020theoretical}
\begin{equation}
\begin{split}
\theta_{pk}=\Bigg\{\frac{-K_1 tr(\bar{\bar{D}}_k)+\sqrt{K_1^2tr^2(\bar{\bar{D}}_k)+4 K_4 \varepsilon_{pk}[2K_3(\bar{\bar{D}}_k:\bar{\bar{D}}_k)+K_2 tr^2(\bar{\bar{D}}_k)]}}{2\varepsilon_{pk} K_4} \Bigg\}^2,
\end{split}
\label{equation11}
\end{equation}
where
\begin{equation}
\begin{split}
K_1=\rho_p[1+2(1+e)\varepsilon_{pk} g_0],
\end{split}
\label{equation11-1}
\end{equation}
\begin{equation}
\begin{split}
K_2=\frac{4}{3\sqrt{\pi}}d_p \rho_p (1+e)\varepsilon_{pk} g_0-\frac{2}{3}K_3,
\end{split}
\label{equation11-2}
\end{equation}
\begin{equation}
\begin{split}
K_3=\frac{d_p \rho_p}{2}\Bigg\{\frac{\sqrt{\pi}}{3(3-e)}\bigg[1+\frac{2}{5}(1+e)(3e-1)\varepsilon_{pk} g_0\bigg]+ \frac{8\varepsilon_{pk}}{5\sqrt{\pi}}g_0(1+e)\Bigg\},
\end{split}
\label{equation11-3}
\end{equation}
\begin{equation}
\begin{split}
K_4=\frac{12(1-e)^2 \rho_p g_0}{d_p \sqrt{\pi}}.
\end{split}
\label{equation11-4}
\end{equation}

The effective shear and bulk viscosities of suspensions can also be defined as $\mu_k=\mu_{gk}\varepsilon_{gk}+\mu_{pk}\varepsilon_{pk}$ and $\lambda_k=\lambda_{gk}\varepsilon_{gk}+\lambda_{pk}\varepsilon_{pk}$ \citep{enwald1996eulerian}, where the granular viscosities $\mu_{pk}$ and $\lambda_{pk}$ can be predicted using kinetic theory of granular flow and/or frictional stress model \citep{srivastava2003analysis,schneiderbauer2012comprehensive}, for example, equation (\ref{d5}). If full granular temperature equations need to be solved to obtain the granular temperatures, significant increase of computational cost is expected. Present study will not touch it, since detailed review has already available \citep{wang2020continuum}. Finally, turbulence models have been used to predict the turbulent shear viscosities \citep{krishna2001using,gao2013novel,gao2015novel}.

\subsubsection{Interphase mass transfer rate}
Very few studies have involved the interphase mass transfer rates between dilute phase and dense phase, which are conceptually in accordance to present study. At the outset it must be stressed that the interphase mass transfer studied here is completely different with the extensive studies of bubble-to-emulsion mass transfer of gases when studying mass transfer and the particle exchange between the bubble wakes (or clouded bubbles) and the emulsion phase when studying the mixing and segregation of particles in bubbling fluidized beds \citep{chiba1977solid,kunii1991fluidization,basesme1992solids,medrano2017determination,wang2022insight}. Moreover, researchers have attempted to develop an interphase mass transfer model using kinetic theory \citep{ozarkar2008kinetic,zhao2021statistical}, unfortunately, these models include parameters that are undetermined. Therefore, these not-ready-to-use models are not touched here.

In the classic two-phase modeling of bubbling fluidized beds, it is customary to assume that the solid volume fraction inside the bubbles is a constant. Furthermore, the gas is assumed to be incompressible in present study. Therefore, the interphase mass transfer rate can be estimated as
\begin{equation}
\begin{split}
\Gamma_f=-\Gamma_d=f_f(\rho_g\varepsilon_{gf}+\rho_p\varepsilon_{pf})\frac{d (\text{ln}V_b)}{d t},
\end{split}
\label{equation11-5}
\end{equation}
where $V_b$ is the volume of a bubble, $\frac{d V_b}{d t}$ is the bubble volume growth rate and $\frac{f_f}{V_b}$ is the number density of bubbles. Of course, only the continuous change of bubbles is eligible to contribute $\frac{d V_b}{d t}$, the bubble growth and/or shrink caused by the coalescence and/or breakup of bubbles can not be accounted here, since they do not change the total volume of involving bubbles. In particular, most of two-phase models available have assumed there is no particles inside bubbles ($\varepsilon_{pf}=0$) \citep{kunii1991fluidization} and no reliable model is available for $\frac{d V_b}{d t}$, therefore, it seems reasonable to assume, at least temporally, $\Gamma_f=0$ in bubbling fluidized beds, as practiced in recent studies \citep{krishna2001using,gao2013novel,gao2015novel,wang2019coupled,wen2022quasi}.

A recent study has shown that although the time-averaged net mass, momentum and energy exchanges are zero in clustered gas-solid flow, instantaneous mass, momentum and energy exchanges between dense phase and dilute phase are extensive \citep{Kong2023Characterizing}, therefore, the interphase mass transfer rate is nonzero in circulating fluidized bed risers at the temporal scale of the time step used in numerical simulations. However, modeling of the interphase mass transfer is sparse. Except for the theoretical analysis using kinetic theory which contains undetermined parameters \citep{ozarkar2008kinetic,zhao2021statistical}. To our best knowledge the only model that could be directly used is due to the work of \citet{hu2017quantifying},
\begin{equation}
\begin{split}
\Gamma_d=\frac{f_d\rho_p}{2d_{cl}}\bigg(3\mid {\bf{u}}_{pf}-{\bf{u}}_{pd}\mid \varepsilon_{pf}-\varepsilon_{pd}\sqrt{{3\theta_{pd}}}\bigg),
\end{split}
\label{a284e}
\end{equation}
where $d_{cl}$ is the diameter of cluster, the granular temperature of dense phase ($\theta_{pd}$) is estimated using the following equation \citep{sangani1996simple,hu2017quantifying}
\begin{equation}
\begin{split}
\frac{\sqrt{\theta_{pd}}}{\mid {\bf{u}}_{pf}-{\bf{u}}_{pd}\mid}=\frac{\sqrt{\pi}R_{diss}}{8\varepsilon_{pd}g_0(1-e)\text{St}}
\bigg\{\bigg[1+\frac{256\varepsilon_{pd}^2(1-e)g_0^2}{15\pi}\frac{\text{St}^2}{R_{diss}^2}\bigg(1+\frac{\pi(1+\frac{5}{8\varepsilon_{pd}g_0})^2}{12}\bigg)\bigg]^{\frac{1}{2}}-1\bigg\},
\end{split}
\label{a284f}
\end{equation}
where $e$ is the restitution of coefficient, $g_0$ is the radial distribution function at contact (equation (\ref{d7})), $\text{St}=\frac{d_p\mid {\bf{u}}_{pf}-{\bf{u}}_{pd}\mid}{9\mu_g}$ is the Stokes number, $R_{diss}$ is a dissipation coefficient characterizing gas-particle interaction. It can be seen that the model has correct limits at two ends: When the solid concentration is approaching zero, the cluster size and $f$ are respectively approaching the particle diameter and zero, hence $\Gamma_d=0$; When the solid concentration is approaching the value at the jamming transition point, the bed is in the packed bed regime, therefore, $\theta_{pd}=0$ and ${\bf{u}}_{pf}={\bf{u}}_{pd}$, hence $\Gamma_d=0$.

\subsubsection{Interfacial force density}
\begin{figure}[!htb]
\centering
\includegraphics[width=0.9\textwidth]{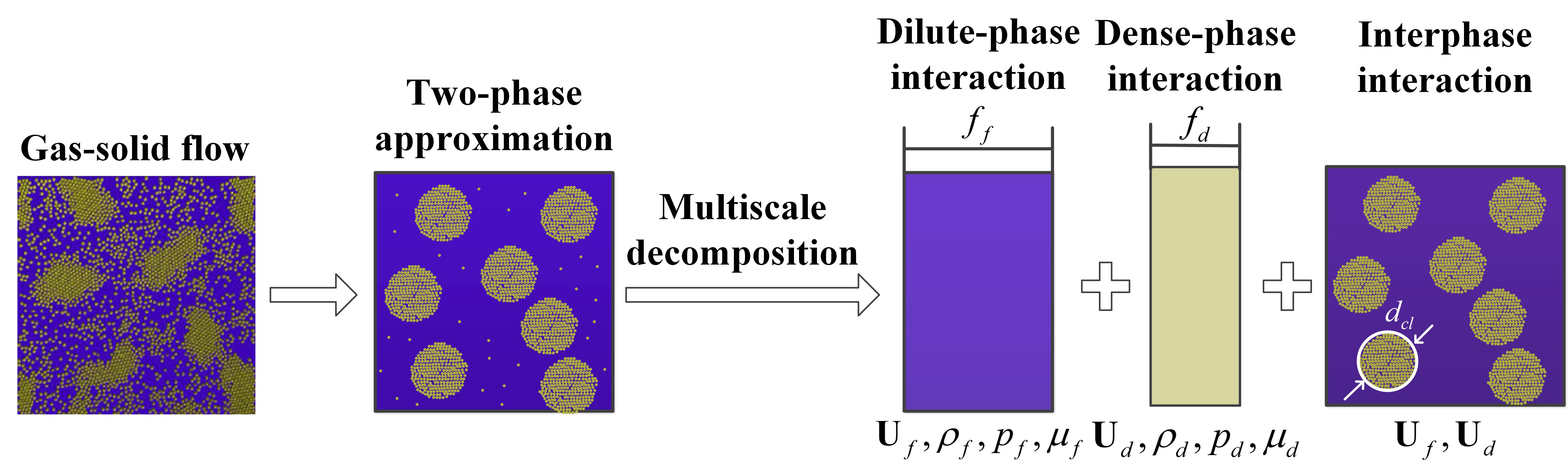}
\caption{Multiscale decomposition for developing the constitutive laws for heterogeneous gas-solid flows}
\label{figconstitutivelaw}
\end{figure}

The interfacial force density ${\bf{M}}_{k}^F$ is critical for the mesoscience-based structural theory. As will be seen below, the models for ${\bf{M}}_{k}^F$ have made full use of two-phase models, the validity of which is normally dependent on fluidization regime, therefore, cluster-based model and bubble-based model are discussed respectively. On the other hand, we emphasize that the two-phase approximation as shown in Figure \ref{figconstitutivelaw} is only needed in the establishment of constitutive laws, no such an approximation is needed when formulating the conservation equations of mesoscience-based structural theory.

The cluster-based constitutive laws for mesoscience-based structural theory can be established using the concept of multiscale decomposition of the EMMS model \citep{li1994particle}, as shown in Figure \ref{figconstitutivelaw}. It is assumed that the heterogeneous gas-solid flow with dynamic, amorphous clusters can be approximated using a dilute-dense two-phase structure, where the dense phase takes the form of monodisperse, spherical clusters, the individual particles in the dilute phase are distributed homogeneously, and the clusters having a  diameter $d_{cl}$ are also homogeneously distributed in the dilute phase. Multiscale decomposition is then carried out to further approximate the dilute-dense two-phase structure as the weighted sum of three homogeneous subsystems, the gas-solid interactions are then approximated as the interactions within the dilute phase and the dense phase, and the interphase interaction between the dilute phase and the dense phase. Since the mesoscience-based structural theory treats the dilute phase and the dense phase as the two interpenetrating continua, the interactions within the dilute phase and the dense phase are represented by the effective stress tensors. However, the interphase interaction between the dilute phase and the dense phase is represented by the interfacial force density studied here, which is closed as follows \citep{gibilaro1985generalized,wang2012emms}
\begin{equation}
\begin{split}
{\bf{M}}_{f}^F=-{\bf{M}}_{d}^F=-\Bigg(\frac{17.3}{\text{Re}}+0.336 \Bigg)\frac{\rho_f f_df_f^{-1.8}\mid{\bf{u}}_f-{\bf{u}}_d\mid}{d_{cl}}({\bf{u}}_f-{\bf{u}}_d),
\end{split}
\label{a133a}
\end{equation}
where the Reynolds number is defined as
\begin{equation}
\begin{split}
\text{Re}=\frac{f_f\rho_fd_{cl}\mid{\bf{u}}_f-{\bf{u}}_d\mid}{\mu_f}.
\end{split}
\label{a133aa}
\end{equation}
Clearly, an expression for $d_{cl}$ needs to be provided and there are many empirical correlations available in literature as summarized previously \citep{wang2020continuum}. Unfortunately, those correlations are normally a function of mean solid concentration, but it is not a state variable in the mesoscience-based structural theory, therefore, they are not ready-to-use (but possibly be transformed using $\varepsilon_s=f_f \varepsilon_{pf}+f_d \varepsilon_{pd}$ if $\varepsilon_{pf}$ and $\varepsilon_{pd}$ are known). Relevant studies \citep{wang2012emms,zhou2014three,zhao2015cfd,zhao2016generalized} have used the EMMS model \citep{li1994particle} to predict the diameter of cluster which is expressed as $d_{cl}=f(f_d)$ or simply assume it as a constant \citep{gao2015novel}. In numerical simulations, with  known value of $f_d$ of a computational grid, the corresponding $d_{cl}$ can be determined.

The bubble-based interfacial force density has only included the drag force in the pioneering study of \citet{krishna2001using}
\begin{equation}
\begin{split}
{\bf{M}}_{f}^F=-\frac{3}{4}\frac{\rho_d f_f f_d}{d_{b}}C_D\mid{\bf{u}}_f-{\bf{u}}_d\mid({\bf{u}}_f-{\bf{u}}_d),
\end{split}
\label{a134a}
\end{equation}
where the interphase drag coefficient is
\begin{equation}
\begin{split}
C_D=\frac{4}{3}\frac{\rho_d-\rho_f}{f_d\rho_d}gd_b\frac{1}{V_b^2},
\end{split}
\label{a134b}
\end{equation}
where $g$ is the gravitational acceleration, the bubble size $d_b$ and the bubble rising velocity $V_b$ are calculated respectively from
\begin{equation}
\begin{split}
d_b=0.204(U-U_{d})^{0.412},
\end{split}
\label{a134c}
\end{equation}
\begin{equation}
\begin{split}
V_b=0.71\sqrt{gd_b}(\text{AF})(\text{SF}),
\end{split}
\label{a134d}
\end{equation}
where $U$ is the operating superficial gas velocity and $U_d$ is the superficial gas velocity inside the emulsion phase that is assumed to be zero in their CFD simulations, AF represents the bubble wake acceleration effect
\begin{equation}
\begin{split}
\text{AF}=1.64+2.7722(U-U_{d}),
\end{split}
\label{a134e}
\end{equation}
and SF is the scale factor that represents the effect of reactor diameter
\begin{equation}
\begin{split}
&\text{SF}=\left\{\begin{array}{ll}
1 & \text{for} \ \frac{d_b}{D_T}\leq 0.125 \\
1.13\text{exp}\big(-\frac{d_b}{D_T}\big)& \text{for} \  0.125<\frac{d_b}{D_T}\leq 0.6\\
0.496\sqrt{\frac{D_T}{d_b}} & \text{for} \ \frac{d_b}{D_T}>0.6
\end{array}\right. \\
\end{split}
\label{a134f}
\end{equation}
where $D_T$ is the diameter of fluidized bed reactor. On the other hand, \citet{gao2013novel} has used different bubble size correlation and drag coefficient, they are
\begin{equation}
\begin{split}
d_b=\bigg(-\Upsilon+\sqrt{\Upsilon^2+4d_{bm}/D_T}\bigg)^2\frac{D_T}{4},\  \text{with} \
d_{bm}=1.49g^{-2}\big[ (U-U_{mf})\pi D_T^2\big]^{0.4}, \ \Upsilon=0.0256\sqrt{D_T/g}/U_{mf},
\end{split}
\label{a134g}
\end{equation}
\begin{equation}
\begin{split}
&C_D=\left\{\begin{array}{ll}
\frac{24}{\text{Re}}\big(1+0.15\text{Re}^{0.687}\big) +\frac{0.413}{1+16.3\text{Re}^{-1.09}}& \text{Re}\leq1000 \\
0.44 & \text{Re}>1000
\end{array}\right. \\
\end{split}
\label{a134h}
\end{equation}
where $\text{Re}=\frac{d_b\mid{\bf{u}}_f-{\bf{u}}_d\mid\rho_d}{\mu_d}$ and $U_{mf}$ is the minimum fluidization velocity.
Clearly, both correlations assume a constant bubble size along the height of the reactor, which is an important limitation. Furthermore, there are many other correlations for predicting bubble size and bubble velocity \citep{karimipour2011critical}, which predict significant differences of bubble behaviour. Those empirical correlations could in principle also be used in the mesoscience-based structural theory, for example, the correlation of \citet{cai1994quantitative} could be used to predict a bed-height-dependent bubble size.
Since the emulsion phase is treated as a continuum, the behaviour of bubbles in gas-solid fluidized beds is analogous to that of bubbly flow in gas-liquid systems, therefore, more interfacial forces may be considered \citep{bokkers2006modelling,gao2013novel,gao2015novel}, including the lift force, the added mass force, the lubrication wall force and the turbulent dispersion force.
We finally note that there are few studies that use population balance equation to predict the spatiotemporal variation of the diameter of clusters or bubbles in gas-solid systems \citep{zhao2016generalized,wang2019coupled,hu2020cfd,hu2021cfd,wang2022computational,wen2022development,hu20233d}, those results could be coupled with the mesoscience-based structural theory to predict the needed cluster/bubble diameter.

\section{Numerical validation}
The mesoscience-based structural theory, as an alternative to the popular two-fluid model, is developed for simulating the heterogeneous gas-solid flows by treating the dilute and dense phases as the two interpenetrating continua, which correspond to the physical realizations of the two dominant mechanisms existing in the system. However, as shown in the preceding sections, it is still not a closed set of equations with the conservation equations and the basic assumption developed and the constitutive laws summarized in Section 2, since it is mathematically formulated as partial differential equations constrained dynamic optimization. In this section, the theory is preliminarily validated by assuming the densities of dilute and dense phases (or the voidage inside the dilute and dense phases) are constants. This assumption results in a model that is essentially the same as the two-fluid model for gas-droplet flows in the case of simulating gas-solid flow in circulating fluidized bed risers or gas-liquid bubbly flows when simulating gas-solid flow in bubbling fluidized beds. Hence no dynamic optimization is needed and the mass and momentum conservation equations are sufficient to describe the dynamics of heterogeneous gas-solid flows. Moreover, the simplified model used in CFD simulations are essentially same as the model of \citet{zhou2014three}, the only main difference lies at that present article systematically studies the effects of the key input parameter (the correlation of cluster diameter) on the simulation results. Finally, the grid dependence has been studied in our previous simulations \citep{zhou2014three}, it has been shown that the simulation results of mesoscience-based structural model are insensitive to the grid size, therefore, no grid dependence study is repeated here.

The velocities at the velocity-inlet of gas phase and the velocity-inlet of clustered dense phase are determined by specific operation conditions (the superficial gas velocity and the solid circulation flux). The boundary conditions for the dense-phase and dilute-phase velocities are both no-slip at the walls. The boundary condition for pressure at the outlet is equal to the atmosphere pressure and is zero-gradient at the walls. Initially, there is no particle in the riser. The total physical time for our simulations to run is 30 s, with the adaptive time step according to the courant number $Co$ which is defined as $Co=\frac{\triangle t \mid u_r \mid}{\triangle x}$ where $\mid u_r \mid$ is the magnitude of the interphase slip velocity, $\triangle t$ is the time step and $\triangle x$ is the grid size. The maximum of $Co$ is 0.9. The time-averaged statistics are carried out at 15-30 s when the flow is fully developed and reached the statistically steady state. The finite volume method is used to discretize the governing equations, with a Total Variation Diminishing (TVD) differencing scheme (limited linear 1) for the convective term, the first-order backward Euler scheme for the transient term and the second-order central differencing scheme for the diffusion term. The PIMPLE algorithm available in OpenFOAM$^\circledR$ is used to solve the velocity-pressure coupling problem.

\begin{figure}[!htb]
\centering
\includegraphics[width=0.7\textwidth]{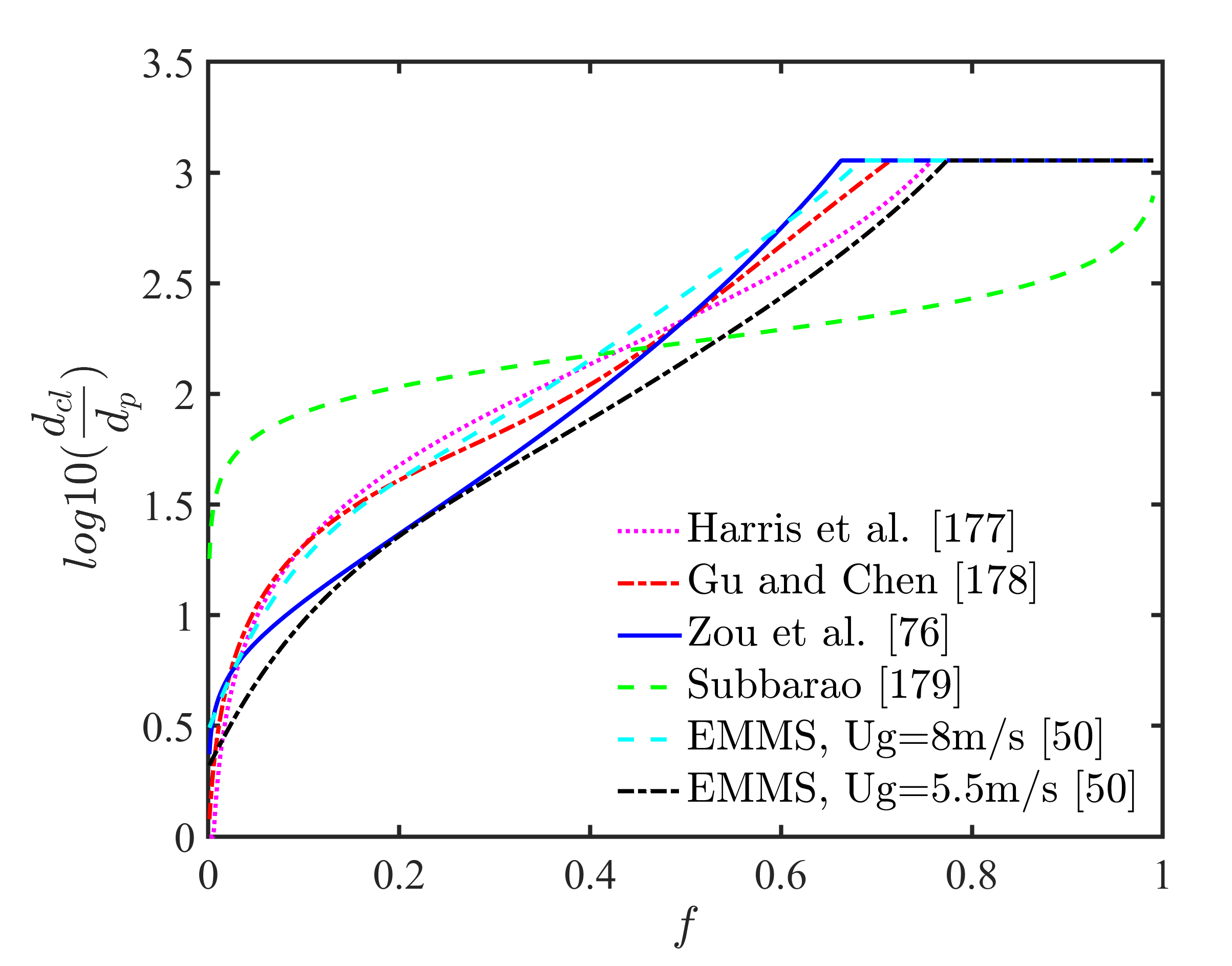}
\caption{Cluster size as a function of the fraction of dense phase $f$. Note that different operational conditions result in different cluster size functions, therefore, there are two lines predicted as the EMMS model.}
\label{dcl_comparison_Ug8_limitD}
\end{figure}

The hydrodynamics of a high-density circulating fluidized bed riser, which corresponds to the experimental facility of \citet{parssinen2001axial,parssinen2001particle}, is simulated using the simplified mesoscience-based structural theory. The simulation parameters including the voidage inside the dilute and dense phases are summarized in Table \ref{tab1}, which are in accordance to the experimental studies \citep{parssinen2001axial,parssinen2001particle}. Following our previous studies \citep{wang2012emms,zhou2014three}, we have assumed  $\varepsilon_{gf}=1$ and $\varepsilon_{gd}=0.5$, then we have $f=2\varepsilon_{s}$.
The effective viscosities of the dilute phase and the dense phase are calculated using equation (\ref{d1}), the interphase mass transfer rates $\Gamma_f$ and $\Gamma_d$ are assumed to be zero following previous studies \citep{krishna2001using,wang2012emms,gao2013novel}, and the interfacial force density is calculated using equation (\ref{a133a}), which essentially means that the dense phase takes the form of clusters and the clusters are homogeneously distributed in the dilute phase. The cluster size in equation (\ref{a133a}) is either predicted theoretically by the EMMS model \citep{li1988method,li1994particle} or determined by one of the empirical correlations summarized in Table \ref{tab2}.
The original publications have not reported the validation range of their empirical correlations of cluster size, on the other hand, all correlations have included the fluidization of typical FCC particles that is also the type of particles used in present study. Therefore, we have assumed that those empirical correlations can be applied in present study. Furthermore, in the EMMS model, the cluster size is a function of operating condition (the superficial gas velocity and the solid circulation flux) and the physical properties of gas (density and viscosity) and particles (density and diameter). The theoretically predicted cluster size is then correlated as a function of the fraction of dense phase ($f$) and fed into the CFD simulations.
Figure \ref{dcl_comparison_Ug8_limitD} shows the cluster size used in the numerical simulations, where the cluster size is assumed to be no larger than the riser diameter. This is the reason why the cluster size with a large value of $f$, except for the correlation of \citet{subbarao2010model}, is a constant. It can be expected that the observed significant differences of cluster size will result in significantly different results of numerical simulations, since the model for the interphase force density is critical for continuum simulations \citep{wang2020continuum}.

\begin{longtable}[H]{lcc}
    \caption{Summary of parameters used in simulations \citep{parssinen2001axial} \label{tab1}}
    \\\hline
    \textbf{Properties} & \textbf{Value} & \quad
    \\\hline
    Riser diameter $D(m)$                   & 0.76 \\
    Riser height $H(m)$                     & 10.0 \\
    Particle diameter $d_p(\mu m)$              & 67 \\
    Particle density $\rho_p(kg/m^3)$       & 1500 \\
    Gas density $\rho_g(kg/m^3)$            & 1.2 \\
    Gas viscosity $\mu_g(kg/m\cdot s)$      & $1.78\times10^{-5}$ \\
    Superficial gas velocity $U_g(m/s)$     & 8.0 & 5.5 \\
    Solids flux $G_s(kg/(m^2s))$      & 550 & 300 \\
    Voidage inside dilute phase $\varepsilon_{gf}$ & 1\\
    Voidage inside dense phase $\varepsilon_{gd}$ & 0.5\\
    Grid number     & 330165 \\
    \hline
\end{longtable}

\begin{longtable}[H]{lcc}
    \caption{Summary of empirical correlations of cluster size \label{tab2}}
    \\\hline
    \textbf{Correlations} & \textbf{Authors} & \quad
    \\\hline
    $d_{cl} = \frac{\varepsilon_s}{40.8 - 94.5 \varepsilon_s}$                   & \citet{harris2002prediction} \\
    $d_{cl} = d_p + (0.027 - 10d_p)\varepsilon_s + 32 \varepsilon_s ^6$                     &\citet{gu1998model} \\
    $d_{cl} = d_p + (\frac{1-\varepsilon_g}{\varepsilon_g - \varepsilon_{gd}})^{\frac{1}{3}} \frac{2u_t^2}{g} (1 + \frac{u_t^2}{(0.35\sqrt{gD})^2})^{-1} $              &\citet{subbarao2010model} $^a$\\
    $\frac{ d_{cl} } {d_p} = 1+1.8543[\frac{\varepsilon_g^{-1.5}\varepsilon_s^{0.25}}{(\varepsilon_g - \varepsilon_{mf})^{2.41}}]^{1.3889}$       &\citet{zou1994cluster} $^b$\\
    \hline
\footnotesize{$^a$$u_t$ is the terminal velocity of particles, $D$ is the diameter of riser.}\\
\footnotesize{$^b$$\varepsilon_{mf}$ is the minimum fluidization voidage.}
\end{longtable}

\begin{figure}[!htb]
\centering
\subfigure[]{
\includegraphics[width=0.45\textwidth]{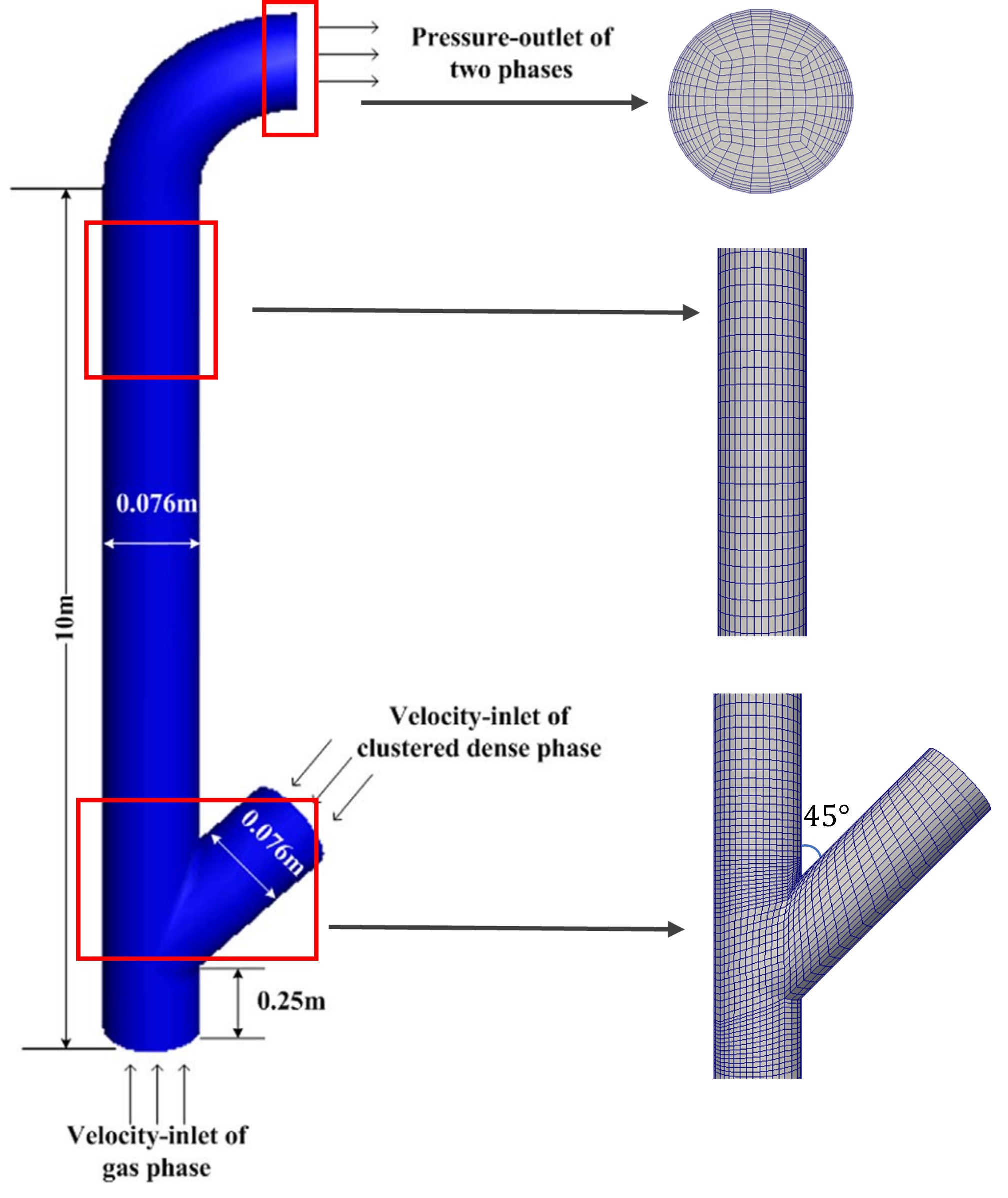}
}
\subfigure[]{
\includegraphics[width=0.52\textwidth]{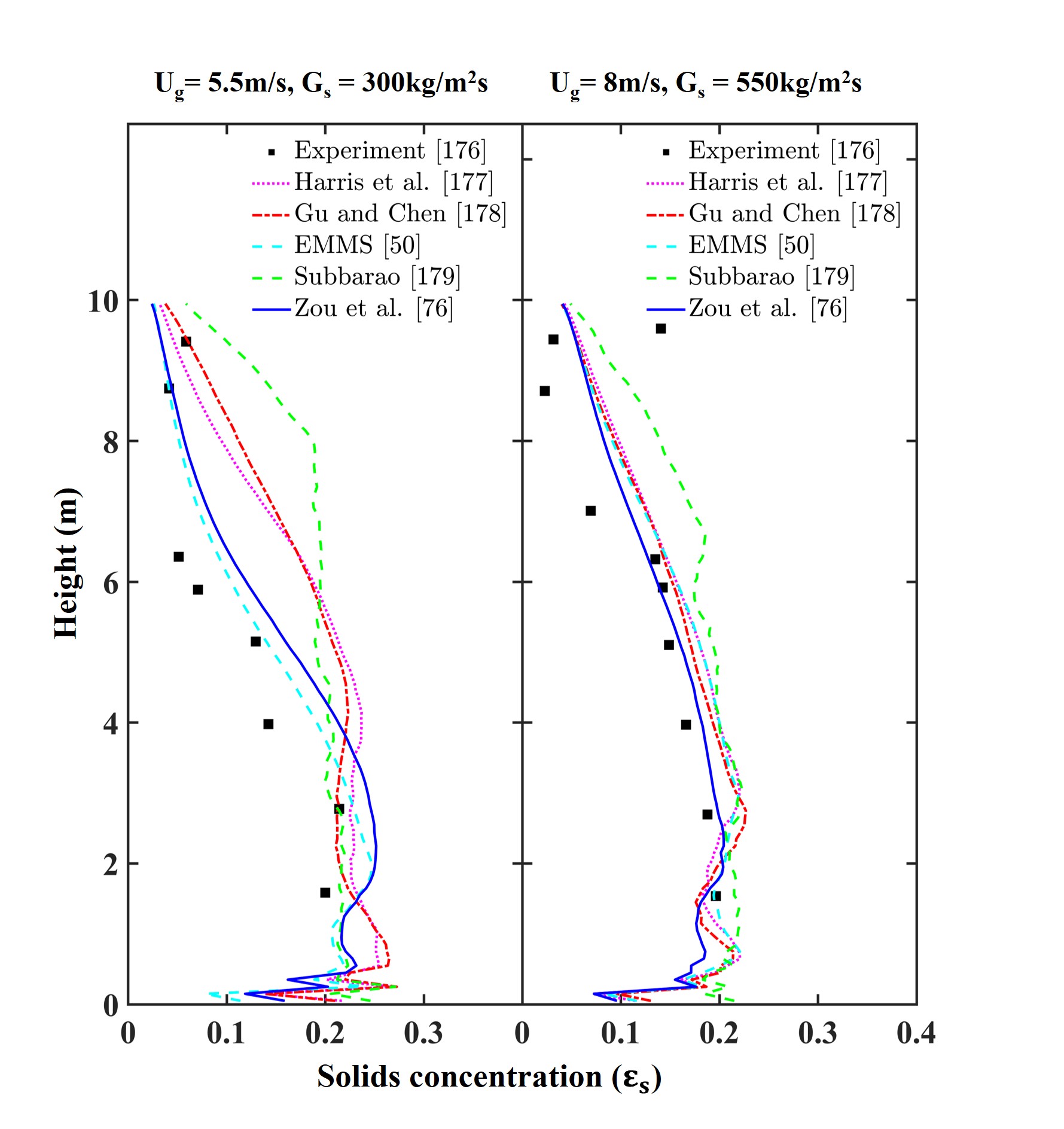}
}
\caption{Not-at-scale schematic of the three dimensional riser and used grids (a), and the comparison of simulated and experimentally measured axial solid concentration profiles (b). In present and hereafter figures, the experimental data are provided by \citet{parssinen2001axial}; the legends of \citet{harris2002prediction}, \citet{gu1998model}, EMMS \citep{li1994particle}, \citet{subbarao2010model} and \citet{zou1994cluster} means that the CFD results are obtained using the simplified mesoscience-based structural model with the corresponding correlation of cluster size of those studies.}
\label{Axialprofiles}
\end{figure}

Figure \ref{Axialprofiles}(a) shows the geometry of the simulated riser with 76mm-ID and 10m height and the used boundary conditions, which is discretized using uniform hexahedral grids. According to a previous grid-independence study on the same CFB riser \citep{zhou2014three}, a grid number of 330165 in total is adequate to obtain grid-size-independent results. therefore, grid sensitivity test haven't been carried out in present study.
Figure \ref{Axialprofiles}(b) shows the time-averaged comparison of simulated and experimental axial solid volume fraction profiles \citep{parssinen2001axial,parssinen2001particle}. It is evident that the empirical correlation of \citet{zou1994cluster} and the cluster size predicted by the EMMS model \citep{li1994particle} fit the experimental data better than others in both cases, while results obtained by \citet{subbarao2010model} are far from the experimental data, especially in the upper part of the riser. Besides, empirical correlation of \citet{harris2002prediction} and \citet{gu1998model} performed almost identical. Different operational conditions result in different cluster size functions, and the best agreement with the experimental data is predicted by the EMMS model in the case of $U_g=5.5m/s$, $G_s=300kg/(m^2s)$ but the closest results to the experimental data in the case of $U_g=8m/s$, $G_s=550kg/(m^2s)$ is obtained using the correlation of \citet{zou1994cluster}.

\begin{figure}[!htb]
\centering
\subfigure[]{
\includegraphics[width=0.48\textwidth]{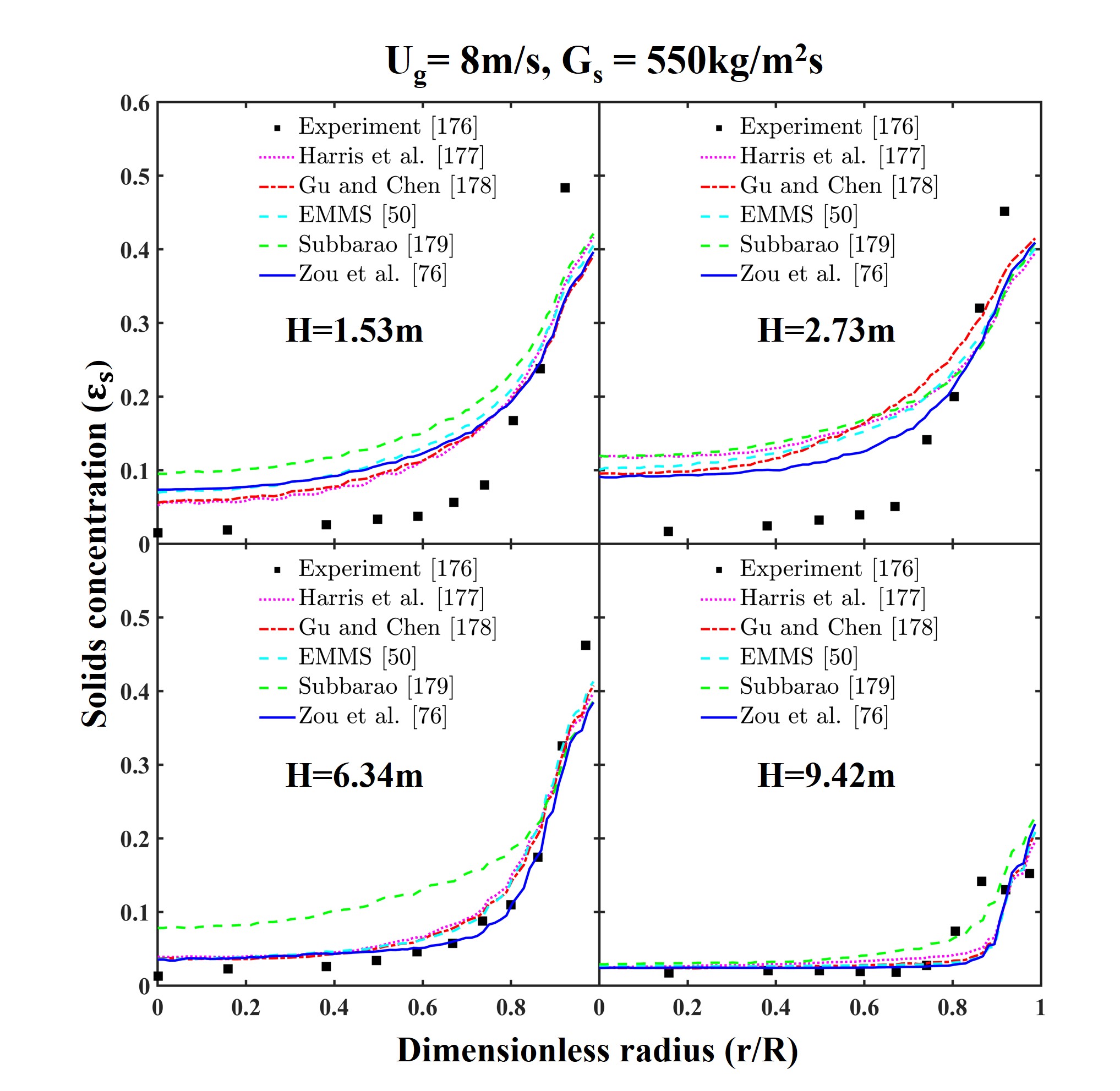}
}
\subfigure[]{
\includegraphics[width=0.48\textwidth]{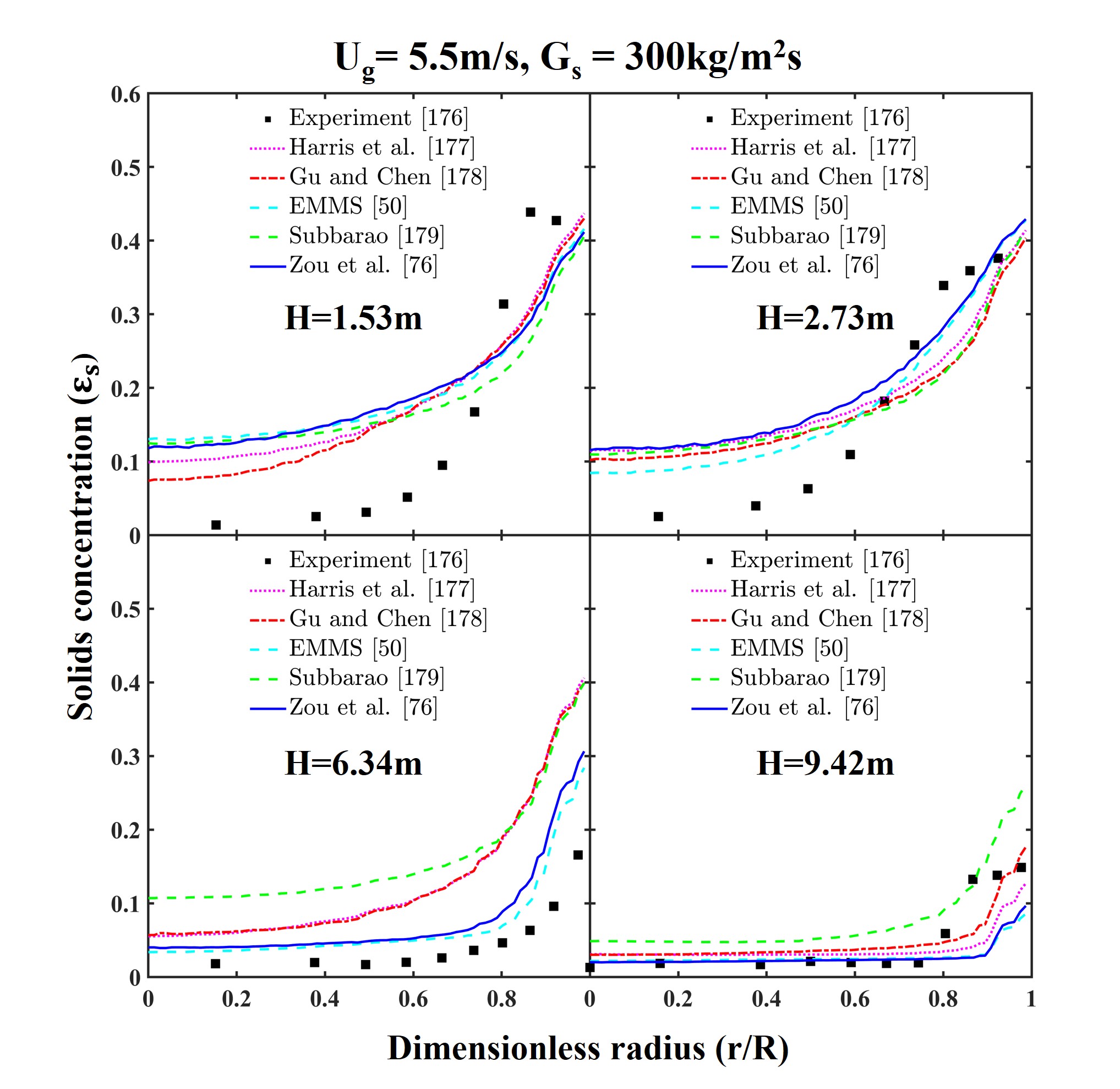}
}
\caption{Comparison of simulated and experimental radial solid volume fraction profiles at four different heights of the bed.}
\label{radialsolidsprofiles}
\end{figure}

\begin{figure}[!htb]
\centering
\subfigure[]{
\includegraphics[width=0.48\textwidth]{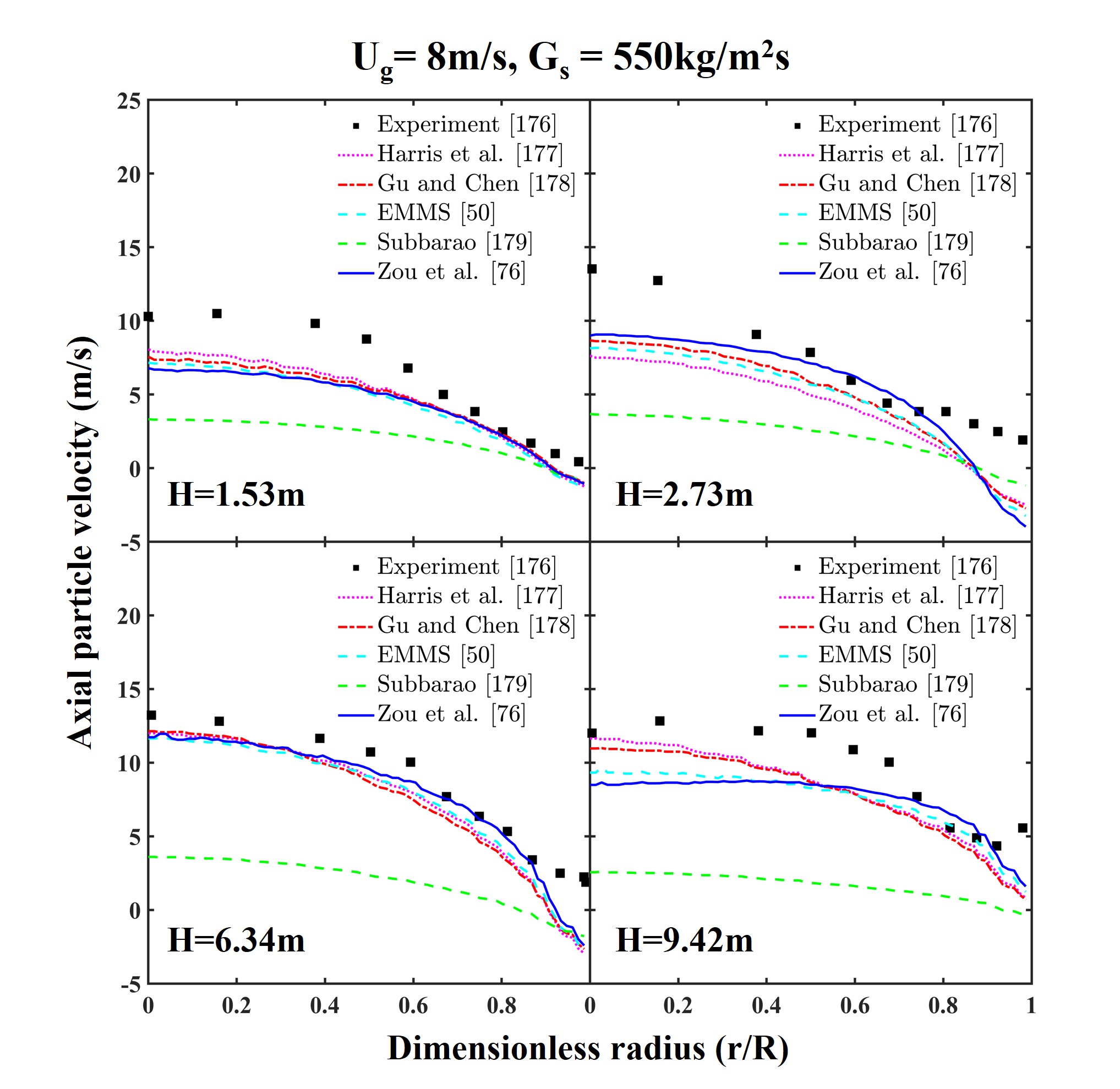}
}
\subfigure[]{
\includegraphics[width=0.48\textwidth]{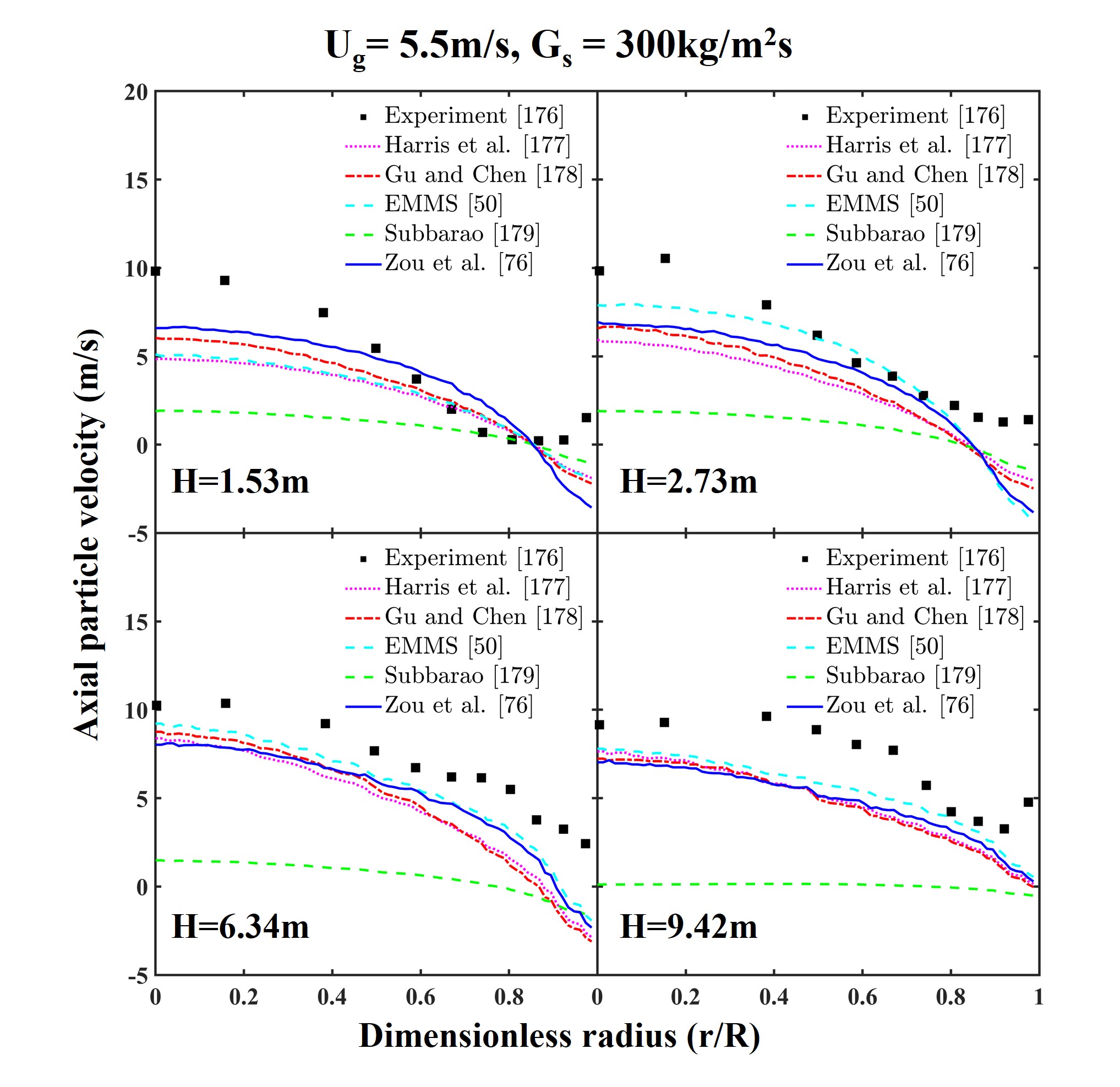}
}
\caption{Comparison of simulated and experimental radial particle velocity profiles at four different heights of the bed.}
\label{radialusprofiles}
\end{figure}

Figure \ref{radialsolidsprofiles} and  Figure \ref{radialusprofiles} show respectively the comparison of simulated and experimental radial solid volume fraction profiles and particle velocity profiles at four different heights of the bed.
It can be seen that in general the agreement with the experimental data is satisfactory except for using the correlation of \citet{subbarao2010model}. The significant deviation from the experimental data can be attributed to the over-prediction of cluster size as shown in Figure  \ref{dcl_comparison_Ug8_limitD}. Due to the significantly over-predicted cluster size, the interphase slip velocity is significantly over-predicted and therefore the predicted particle velocity is too small as shown in Figure \ref{radialusprofiles}. Furthermore, larger cluster size results in smaller interphase drag force, therefore, the interphase interaction is ineffective to carry the particles out of the riser, which results in the significantly over-predicted axial and radial solid concentrations as shown in Figures \ref{Axialprofiles} and \ref{radialsolidsprofiles}. Following those analysis, it may conclude that all other correlations and the EMMS model also over-predict the cluster size at the relatively small solid concentrations, since the solid concentrations and particle velocities at the central part of the riser are over-predicted and under-predicted respectively. Those findings highlight the need of better models/correlations for predicting the cluster size.

\section{Conclusion}
In response to the fact that the phenomenon of the coexistence of dilute and dense phases and its spatiotemporal evolution in heterogeneous gas-solid two-phase flows is due to the compromise in competition between the gas-dominant mechanism and the particle-dominant mechanism, a  mesoscience-based structural theory is proposed by describing the physical states corresponding to the realization of the two dominant mechanisms as the interpenetrating continua, finding the microscale governing equations of gas-solid flow, defining a dominant mechanism indicator function, and finally, performing ensemble averaging to obtain the macroscopic governing equations. It is found that the mesoscience-based structural theory can be mathematically formulated as partial differential equations (PDE) constrained dynamic optimization, which differs significantly to the popular two-fluid model that are formulated only using PDE (mass, momentum and energy conservation equations). On the other hand, the mesoscience-based structural theory may be regarded as an extension of the static and global version of EMMS model that is mathematically formulated as algebraic equations constrained optimization \citep{li1988method,li1994particle} to a transient and local version, or else be regarded as a much more rigorous formulation of EMMS-based two-fluid model using mesoscience \citep{wang2012emms}. Finally, a simplified version of mesoscience-based structural theory was numerically solved to simulate the gas-solid flows in a high-density circulating fluidized bed riser, it was shown that the hydrodynamics can be predicted reasonably well, thus offering a preliminary validation of mesoscience-based structural theory. In the future, it will be interesting to extract the cluster size from fine grid simulations, CFD-DEM simulations or particle-resolved direction numerical simulations using artificial intelligence and/or standard data-fitting, as has already been attempted recently \citep{kong2022discrete,Kong2023Characterizing} and simultaneously to consider the effects of dynamic evolution of clusters in CFD simulations. Furthermore, the mesoscience-based structural theory (or the partial differential equations constrained dynamic optimization) will be directly solved using for example the adjoint method \citep{he2020dafoam} or the physics-informed machine learning method \citep{raissi2019physics,karniadakis2021physics}.

\section*{Acknowledgement}
We thank Professor Jinghai Li at Institute of Process Engineering, Chinese Academy of Sciences for supervision and suggestions on the drafts. This study is financially supported by the National Natural Science Foundation of China (11988102, 21978295), the Strategic Priority Research Program of the Chinese Academy of Sciences (XDA29040200) and the Innovation Academy for Green Manufacture, Chinese Academy of Sciences (IAGM-2019-A13).

\section*{Appendix A: Two-fluid model}
Two-fluid model has been extensively used to study the spatiotemporal dynamics of gas-solid flows \citep{gidaspow1994multiphase,wang2020continuum}. In order to highlight and compare the differences between two-fluid model and present study, the main features of two-fluid model are summarized here.
Two-fluid model uses two sets of Navier-Stokes equations to describe the gas and particle fluid flows \citep{anderson1967fluid,gidaspow1994multiphase}, which is an instinct extension of the Navier-Stokes equations for single-phase molecular flows. The conservation equations are easy to formulate, provided that the continuum assumption is valid.
The mass conservation equations for gas and solid phases are \citep{anderson1967fluid}
\begin{equation}
\begin{split}
&\frac{\partial (\varepsilon_g\rho_g)}{\partial t}
+\nabla\cdot(\varepsilon_g\rho_g{\bf{u}}_g)=0
\end{split}
\label{a1}
\end{equation}
and
\begin{equation}
\begin{split}
&\frac{\partial (\varepsilon_p\rho_p)}{\partial t}
+\nabla\cdot(\varepsilon_p\rho_p{\bf{u}}_p)=0,
\end{split}
\label{a2}
\end{equation}
where $\varepsilon$, $\rho$ and ${\bf{u}}$ are the volume fraction, the density and the velocity of gas phase and solid phase. Furthermore, those equations should satisfy the geometrical constraint, that is,
\begin{equation}
\begin{split}
\varepsilon_p+\varepsilon_g=1.
\end{split}
\label{a3}
\end{equation}
The gas and solid momentum conservation equations are respectively \citep{gidaspow1994multiphase}
\begin{equation}
\begin{split}
\frac{\partial(\varepsilon_g\rho_g{\bf{u}}_g)}
{\partial t}
+\nabla\cdot(\varepsilon_g\rho_g{\bf{u}}_g{\bf{u}}_g)
=-\varepsilon_g\nabla p_g+\nabla\cdot(\varepsilon_g{\bm{\tau}}_g)
+\varepsilon_g\rho_g{\bf{g}}-{\bf{F}}_{drag}
\end{split}
\label{a13}
\end{equation}
and
\begin{equation}
\begin{split}
\frac{\partial(\varepsilon_p\rho_p{\bf{u}}_p)}
{\partial t}
+\nabla\cdot(\varepsilon_p\rho_p{\bf{u}}_p{\bf{u}}_p)
=-\varepsilon_p\nabla p_g+\nabla\cdot(\varepsilon_p{\bf{T}}_p)
+\varepsilon_p\rho_p{\bf{g}}+{\bf{F}}_{drag},
\end{split}
\label{a14}
\end{equation}
where $\bf{g}$ is the gravitational acceleration and $p_g$ is the pressure of gas. It can be seen that there are nine state variables $\varepsilon_g$, $\varepsilon_p$, $p_g$, ${\bf{u}}_g$ and ${\bf{u}}_p$ (each velocity vector has three components) and the number of conservation equations are also nine (equations (\ref{a1})-(\ref{a14})), therefore, once the constitutive relationships for the effective shear stress tensor of gas phase ${\bm{\tau}}_g$ which is a sum of laminar and turbulent stress tensors, the particle phase stress tensor ${\bf{T}}_p$ and the interphase drag force ${\bf{F}}_{drag}$ as functions of some of the nine state variables are provided, the equation set is closed. Note that (i) we have assumed that $\rho_p$ and $\rho_g$ are constants for the purpose of simplicity. If the compressibility of gas needs to be considered, then the equation of state should be added; and (ii) if the constitutive relationship of ${\bf{T}}_p$ is provided by kinetic theory of granular flow, then the granular temperature $\Theta_p$ is the additional state variable and the granular temperature equation is the additional conservation equation \citep{gidaspow1994multiphase}. Therefore, it is crystal clear that {the conservation equations or dynamic equations alone are sufficient to describe the hydrodynamics of gas-solid flows in two-fluid model since the number of state variables is always equal to the number of dynamic equations, although its success or failure critically depends on the constitutive relationships used}.
%This fact is strongly supported by the state-of-the-art two-fluid modeling of gas-solid flows summarized in Figure \ref{figcontinuummodeling}, using TFM-based DNS, filtered method, two-fluid turbulence model or two-fluid model with structural effect.

If two-fluid model is use to simulate the spatiotemporal dynamics of heterogeneous gas-solid flows, it is necessary to develop models to take the effects of sub-grid-scale heterogeneity on the ${\bm{\tau}}_g$, ${\bf{T}}_p$ and ${\bf{F}}_{drag}$ into account, except that TFM-based DNS is carried out to explicitly resolve the mesoscale heterogeneous structures, which is unfortunately computationally unaffordable when facing the simulations of engineering and industrial scale reactors \citep{wang2009two}. Models and theories that quantify the effects of spatiotemporal evolution of heterogeneous structures on the constitutive relationships are much more difficult to develop than those for homogeneous structures \citep{wang2020continuum}. Therefore, present study has developed an alternative for describing the spatiotemporal dynamics of heterogeneous gas-solid flows.

\section*{Appendix B: Alternative formulation}
In the standard mixture model for fluid-particle suspensions \citep{manninen1996mixture,jamshidi2019closure}, another mass balance equation for each fluid is solve to obtain the solid volume fractions ($\varepsilon_{pk}$), and therefore, the densities of dilute and dense phases ($\rho_k$) via equations (\ref{a146}) and (\ref{a147}). The mass balance equations are
\begin{equation}
\begin{split}
\frac{\partial \varepsilon_{pk}\rho_p}{\partial t} +\nabla \cdot (\varepsilon_{pk}\rho_p{\bf{u}}_k)=-\nabla \cdot [\varepsilon_{pk}\rho_p({\bf{u}}_{pk}-{\bf{u}}_k)].
\end{split}
\label{AB1}
\end{equation}
From equation (\ref{AB1}), it is easy to derive the averaged equation
\begin{equation}
\begin{split}
\frac{\partial \varepsilon_{pk} f_k \rho_p}{\partial t} +\nabla \cdot ( \varepsilon_{pk}f_k\rho_p{\bf{u}}_k)=-\nabla \cdot [\varepsilon_{pk}f_k\rho_p({\bf{u}}_{pk}-{\bf{u}}_k)]+\overline{\varepsilon_{pk}{\rho}_p({\bf{u}}_{pk}-{\bf{u}}_{i,k}) \cdot \nabla X_k},
\end{split}
\label{AB2}
\end{equation}
with the equations we can make sure that the number of state variables are equal to that of equations, but need additional models for closing $\overline{\varepsilon_{pk}{\rho}_p({\bf{u}}_{pk}-{\bf{u}}_{i,k}) \cdot \nabla X_k}$ and ${\bf{u}}_{pk}-{\bf{u}}_k$. In this case, mesoscience-based structural theory is not formulated as the partial differential equations constrained dynamic optimization, since dynamic equations alone are sufficient to describe spatiotemporal dynamics of heterogeneous gas-solid flows. However, it is of paramount importance to realize that this formulation is specific to gas-solid flow, but the formulation of partial differential equations constrained dynamic optimization is general to any other systems where mesoscience is applicable, by describing the physical states corresponding to the realization of dominant mechanisms as the interpenetrating continua, providing the microscale governing equations of studied system, defining a dominant mechanism indicator function, and finally performing ensemble averaging to obtain the macroscopic governing equations.

\section*{References}
\bibliographystyle{elsarticle-num-names}
\biboptions{square,numbers,sort&compress}
\bibliography{G_KT_ref}

%\end{CJK*}
%\end{spacing}
\end{document}